\documentclass[letterpaper,prl,aps,superscriptaddress,twocolumn,floatfix]{revtex4-1}
\usepackage[T1]{fontenc}
\usepackage[latin9]{inputenc}
\usepackage{lmodern}
\usepackage{stix}
\usepackage{amsmath}
\usepackage{microtype,bbm}
\usepackage{graphicx}
\usepackage{booktabs}
\usepackage{times}
\usepackage[scaled=.9]{helvet}
\usepackage[usenames,dvipsnames]{xcolor}

\usepackage{hyperref}
\hypersetup{
	colorlinks,
	allcolors=NavyBlue,
	linktoc=all
}

\usepackage[capitalize]{cleveref}
\crefformat{section}{Sec.~#2#1#3}

\newcommand{\dd}{\mathrm{d}}
\newcommand*\ev[1]{\langle #1 \rangle}
\renewcommand{\Re}{\mathrm{Re}}
\renewcommand{\Im}{\mathrm{Im}}
\newcommand{\cc}{\mathcal C}
\newcommand{\G}{\mathcal G}	
	
\newcommand{\N}{\mathcal N}

\newcommand{\Z}{\mathbb Z}
\renewcommand*\O{\mathcal{O}} 
\newcommand*\dagg{^{\dagger}}
\newcommand*\eps{\varepsilon}
\newcommand{\id}{\mathbbm{1}}

\DeclareMathOperator{\diag}{diag}
\let\Re\relax
\let\Im\relax
\DeclareMathOperator{\Re}{Re}
\DeclareMathOperator{\Im}{Im}
\newcommand*\dens{{\rho}}
\newcommand*\mat[1]{\begin{pmatrix}#1\end{pmatrix}} 
\newcommand*\matr[1]{\mathbf{#1}}

\begin{document}
\title{Quantum-limited directional amplifiers with optomechanics}
\author{Daniel Malz}
\affiliation{Cavendish Laboratory, University of Cambridge, Cambridge CB3 0HE, United Kingdom}
\author{L\'aszl\'o D. T\'oth}
\author{Nathan R. Bernier}
\author{Alexey K. Feofanov}
\author{Tobias J. Kippenberg}
\affiliation{Institute of Physics, \'Ecole Polytechnique F\'ed\'erale de Lausanne, Lausanne 1015, Switzerland}
\author{Andreas Nunnenkamp}
\affiliation{Cavendish Laboratory, University of Cambridge, Cambridge CB3 0HE, United Kingdom}
\date{\today}

\begin{abstract}
  Directional amplifiers are an important resource in quantum information processing,
  as they protect sensitive quantum systems from excess noise.
  Here, we propose an implementation of phase-preserving and phase-sensitive directional amplifiers
  for microwave signals in an electromechanical setup comprising two microwave cavities and two mechanical resonators.
  We show that both can reach their respective quantum limits on added noise. 
  In the reverse direction, they emit thermal noise stemming from the mechanical resonators and we discuss how this noise can be suppressed, a crucial aspect for technological applications.
  The isolation bandwidth in both is of the order of the mechanical linewidth divided by the amplitude gain.
  We derive the bandwidth and gain-bandwidth product for both and find that the phase-sensitive amplifier has an unlimited gain-bandwidth product. 
  Our study represents an important step toward flexible, on-chip integrated nonreciprocal amplifiers of microwave signals.
\end{abstract}

\maketitle

\emph{Introduction.}---Nonreciprocal transmission and amplification of signals are essential in communication and signal processing,
as they protect the signal source from extraneous noise.
Conventional ferrite-based devices rely on magnetic fields and are challenging to integrate in superconducting circuits.
Hence, there exists strong incentive to find more suitable implementations~\cite{Gallo2001,Koch2010,Fang2012a,Sounas2013,Kamal2011,Hafezi2012,Poulton2012,Longhi2013,Fang2013,Wang2013,Abdo2014,Poshakinskiy2017,Sliwa2015,Lecocq2017}.
In the microwave domain, the strong Josephson nonlinearity and parametric pumping can achieve both photon gain and conversion processes,
which have been exploited to realize circulators and directional amplifiers~\cite{Kamal2011,Abdo2013,Sliwa2015,Lecocq2017}.
Another promising platform is optomechanics, where nonreciprocal devices
\cite{Ranzani2014,Ranzani2015,Metelmann2015,Fang2017,Kamal2016,Metelmann2017,Shen2016,Ruesink2016,Xu2016,Bernier2017,Peterson2017,Tian2017},
phase-preserving amplifiers~\cite{Massel2011,Metelmann2014,Nunnenkamp2014,Toth2017,Ockeloen-Korppi2016},
and phase-sensitive amplifiers~\cite{Suh2014,Pirkkalainen2015,Wollman2015,Lei2016} have been proposed and realized.

In recent theoretical work, Ranzani and Aumentado~\cite{Ranzani2014,Ranzani2015} analyzed general conditions for nonreciprocity in parametrically coupled systems, and showed that nonreciprocity arises due to dissipation in ancillary modes and multi-path interference.
Metelmann and Clerk~\cite{Metelmann2015} have shown that any coherent interaction can be made directional by balancing it with a dissipative process.
Indeed, this insight led to a demonstration of nonreciprocity using optomechanics in the optical domain~\cite{Fang2017}, and theoretical investigations into minimal implementations of directional amplifiers~\cite{Kamal2016}.

\begin{figure}[t]
	\centering
	\includegraphics[width=\linewidth]{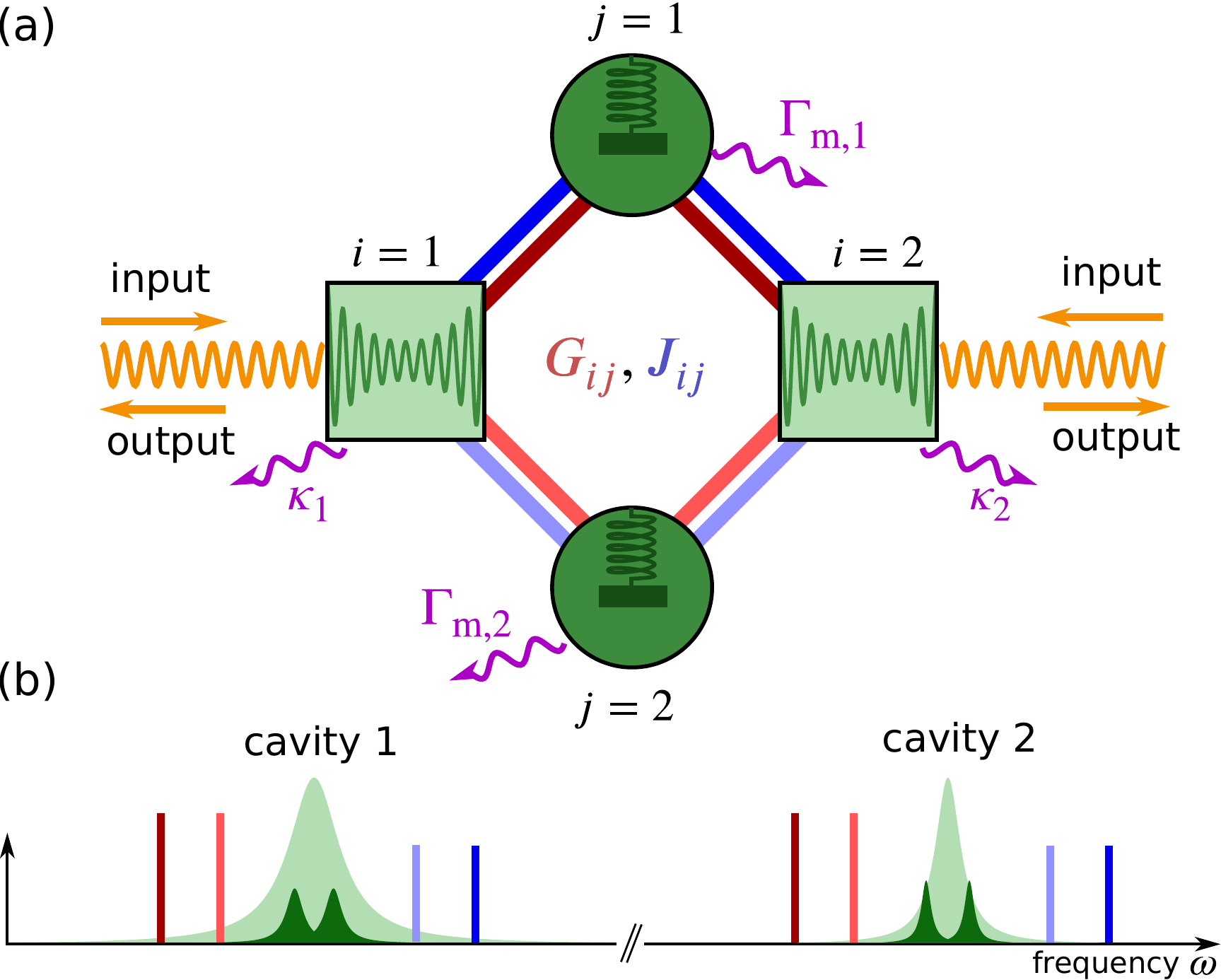}
		\caption{(Color online.) (a) Schematic of all possible interactions in optomechanical plaquette comprising two mechanical resonators (round, dark green) and two cavities (square, light green).
			The cavities [light green Lorentzians in (b)] are driven by up to eight tones, placed close to the mechanical sidebands,
            at frequencies $\omega_{\mathrm c,i}\pm(\Omega_{j}+\delta_j)$, as illustrated in (b),
			which induce hopping and two-mode squeezing interactions ($G_{ij}$, $J_{ij}$),
            denoted in (a) by red and blue lines connecting the modes.
			This leads to the time-independent Hamiltonian~\eqref{eq:Hamiltonian}.
	}
	\label{fig:model}
\end{figure}
While implementing the balance between a \emph{direct} coherent coupling between the cavities and a dissipative interaction is challenging experimentally,
Refs.~\cite{Tian2017,Bernier2017,Peterson2017} have recently studied and demonstrated nonreciprocal transmission
between two cavity modes where two mechanical resonators each mediate both coherent and dissipative coupling.
Here, building on this concept, we propose directional amplifiers using exclusively optomechanical interactions.
Microwave tones on the red and blue sidebands enable so-called beam-splitter and
two-mode squeezing interactions (cf.~\cref{fig:model}), leading to a total of eight controllable terms in the Hamiltonian.
We identify and analyze a simple directional phase-preserving amplifier that uses four tones and
a directional phase-sensitive amplifier using six tones.
While the gain-bandwidth product of the phase-preserving amplifier is limited to the cavity decay rate,
the phase-sensitive amplifier has an unlimited gain-bandwidth product.
We show that both amplifiers can reach their quantum limits of a half and zero added quanta, respectively,
and emit thermal noise from the mechanical resonators in the reverse direction,
a necessary consequence of impedance matching and directionality.
We show how the reverse noise can be reduced through additional sideband cooling without interfering with directionality or amplification.
Our concrete proposal bridges the gap between previous theoretical studies and experimental realization and
therefore represents an important step toward on-chip integrated nonreciprocal amplifiers of microwave signals.

\begin{figure*}[t]
	\centering
	\includegraphics[width=\linewidth]{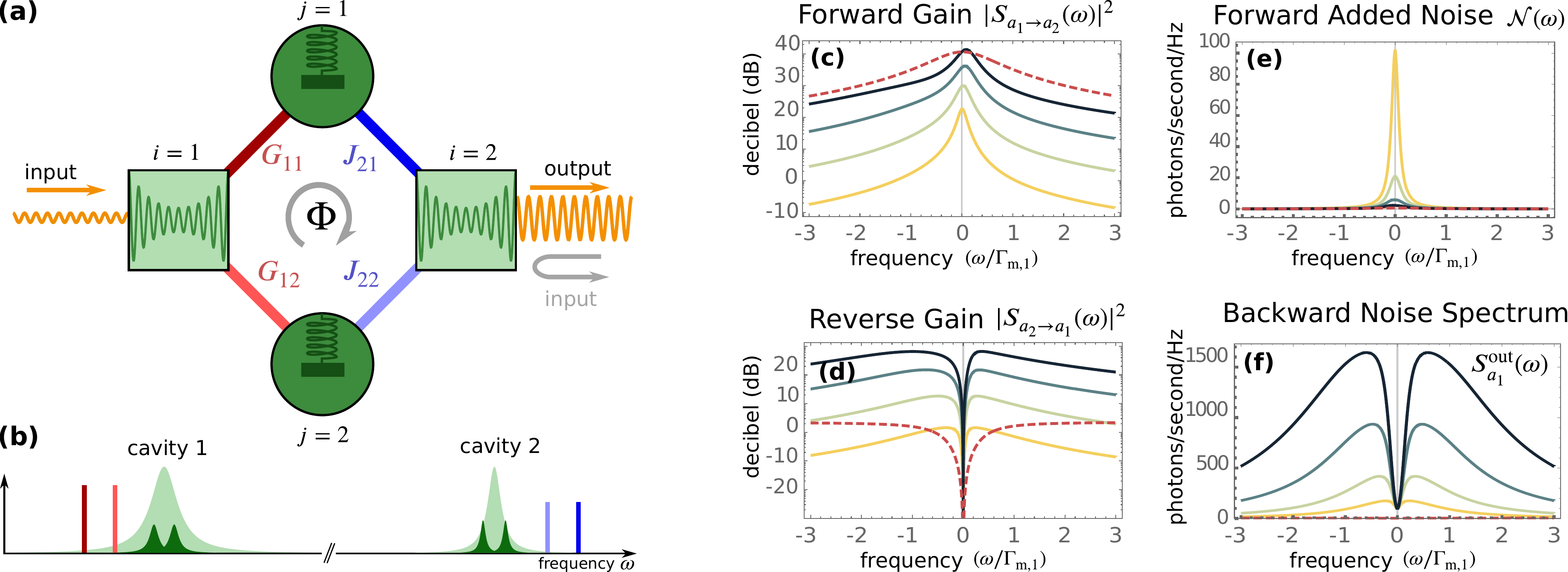}
	\caption{(Color online.) The directional phase-preserving amplifier (DPPA).
		\textbf{(a)} Model. Red hopping interactions are impedance matched, blue provide amplification.
		\textbf{(b)} Pump setup. Cavity 1 (cavity 2) is pumped on the red (blue) sidebands of the mechanical resonators. 
		In \textbf{(c-f)} we plot forward gain, reverse gain [\cref{eq:dppa S}], added noise [\cref{eq:dppa N}], and the output noise fluctuation spectrum of cavity 1, all as functions of frequency in units of $\Gamma_{\mathrm m,1}$,
        for cooperativities $\cc_1=\{1,3,10,30\}$, $\cc_2=\cc_1-0.1\sqrt{\cc_1}$ (yellow to black, or light to dark).
        Parameters are $\kappa_2/\kappa_1=0.7,\Gamma_{\mathrm m,1}/\kappa_1=10^{-2},\Gamma_{\mathrm m,2}/\Gamma_{\mathrm m,1}=0.8$, thermal occupation of the mechanical resonators $n_{\mathrm m,1}=n_{\mathrm m,2}=100$, and cavities $n_{\mathrm c,1}=n_{\mathrm c,2}=0$.
        Depending on parameters, external sideband cooling with an auxiliary mode can achieve $n_{\mathrm m,j}\approx0$, without negatively affecting amplification properties, as discussed below and in the SI.
        The red (dashed) curve in each plot illustrates this case, with 
        $\cc_1=30$ and effective parameters
        $n_{\text{eff},i}=n_{\mathrm m,i}(\Gamma_{\mathrm m,i}/\Gamma_{\text{eff},i}),\Gamma_{\text{eff},i}=50\Gamma_{\mathrm m,i}$.
	}
	\label{fig:dppa}
\end{figure*}

\emph{Model.}---We consider an optomechanical plaquette, comprising two microwave cavities coupled via two mechanical resonators, as shown in \cref{fig:model}(a).
The cavities are driven close to the motional sidebands.
After a standard treatment, which includes linearizing the Hamiltonian, neglecting counterrotating terms,
and going into a rotating frame~\cite{Clerk2010,Aspelmeyer2014}, we arrive at the time-independent Hamiltonian $(\hbar=1)$
\begin{equation}
	H_{\text{sys}}=-\sum_{i=1}^2\delta_ib_i\dagg b_i
	-\sum_{i,j=1}^2a_i\dagg(G_{ij}b_j+J_{ij}b_j\dagg)+\text{H.c.},
	\label{eq:Hamiltonian}
\end{equation}
where $a_i$ ($b_i$) is the annihilation operator for the $i$th cavity mode (mechanical resonator),
$G_{ij}=\alpha_{ij-}g_{0,ij}$ and $J_{ij}=\alpha_{ij+}g_{0,ij}$ are field-enhanced optomechanical coupling strengths,
$\alpha_{ij\pm}$ is the amplitude of the coherent state produced in cavity $i$ due to a pump at frequency $\omega_{\mathrm c,i}\pm(\Omega_{j}+\delta_j$),
and $g_{0,ij}$ are the vacuum optomechanical couplings.
Since the couplings $G_{ij}, J_{ij}$ depend on the pumps, their amplitude and phase can be controlled.
The interactions are represented in Figs.~\ref{fig:model}(a), \ref{fig:dppa}(a), \ref{fig:dpsa}(a) as red ($G_{ij}$) and blue ($J_{ij}$) lines.
Further details can be found in the Supplementary Information (SI), including a discussion about the limits of validity of the rotating-wave approximation.

We describe the system with quantum Langevin equations~\cite{Gardiner1985,gardiner2004quantum,Aspelmeyer2014}.
Neglecting mechanical noise (analyzed later), and eliminating the mechanical modes, we obtain
\begin{equation} 
  	\sum_{j=1}^2[\chi_{\mathrm c,i}^{-1}(\omega)\delta_{ij}+i\matr T_{ij}(\omega)]\vec A_j(\omega)=\sqrt{\kappa_i}\vec A_{i,\text{in}}(\omega),
	\label{eq:2by2}
\end{equation}
where the susceptibility $\chi_{\mathrm c,i}(\omega)=[\kappa_i/2-i\omega]^{-1}$,
$\vec A_{j(,\text{in})}=(a_{j(,\text{in})},a_{j(,\text{in})}\dagg)^T$,
and each $i\matr T_{ij}$ is a 2-by-2 matrix
\begin{multline}
	i\matr T_{ij}(\omega)=\sum_{k=1}^2\sigma_z\left[ \chi_{\mathrm m,k}(\omega)\mat{G_{ik}G_{jk}^*&G_{ik}J_{jk}\\J_{ik}^*G_{jk}^*&J_{ik}^*J_{jk}}\right.\\
	\left.-\chi_{\mathrm m,k}^*(-\omega)\mat{J_{ik}J_{jk}^*&J_{ik}G_{jk}\\G_{ik}^*J_{jk}^*&G_{ik}^*G_{jk}}\right],
	\label{eq:Tij}
\end{multline}
where $\sigma_z=\diag(1,-1)$ and  $\chi_{\mathrm m,i}(\omega)=[\Gamma_{\mathrm m,i}/2-i(\omega+\delta_i)]^{-1}$.
$i\matr T_{ii}$ is akin to a self-energy for mode $A_i$, whereas $iT_{ij}$ for $i\neq j$ is a matrix of coupling strengths between the modes.
Since the interaction is mediated by mechanical resonators, their susceptibility $\chi_{\mathrm m,i}$ appears in the coupling matrix.

Using the input-output relation
$a_{i,\text{out}}=a_{i,\text{in}}-\sqrt{\kappa_i}a_i,$ \cite{Gardiner1985}
the optical scattering matrix is $\matr S_{\text{optical}}(\omega)=\id_4-\matr L\matr \chi(\omega)\matr L$,
where $\matr L=\text{diag}(\sqrt{\kappa_1},\sqrt{\kappa_1},\sqrt{\kappa_2},\sqrt{\kappa_2})$, and
\begin{equation}
	[\matr\chi(\omega)]^{-1}=\mat{\chi_{\mathrm c,1}^{-1}(\omega)\id_2+i\matr T_{11}(\omega)&i\matr T_{12}(\omega) 
	\\i\matr T_{21}(\omega)&\chi_{\mathrm c,2}^{-1}(\omega)\id_2+i\matr T_{22}(\omega)}.
\end{equation}
We say the system is nonreciprocal if the moduli of forward and reverse scattering amplitudes differ, which occurs if $|\matr T_{12}|\neq|\matr T_{21}|$.
Looking for instance at the top left elements $[iT_{12}]_{11},[iT_{21}]_{11}$,
we see that nonreciprocity arises because flipping direction ($1\leftrightarrow2$) conjugates the complex couplings,
but leaves the mechanical susceptibility unchanged.
Nonreciprocity can also be understood in the framework presented in Ref.~\cite{Metelmann2015} (cf.~SI).

\emph{Directional phase-preserving amplifier (DPPA).}---We consider the coupling amplitudes [cf.~\cref{fig:dppa}(a)]
\begin{equation}
	\begin{aligned}
      	\matr G&=\frac12\mat{e^{\frac{i\Phi}{2}}\sqrt{\cc_1\Gamma_{\mathrm m,1}\kappa_1}&e^{-\frac{i\Phi}{2}}\sqrt{\cc_1\Gamma_{\mathrm m,2}\kappa_1}\\0&0},\\
      	\matr J&=\frac12\mat{0&0\\\sqrt{\cc_2\Gamma_{\mathrm m,1}\kappa_2}&\sqrt{\cc_2\Gamma_{\mathrm m,2}\kappa_2}},
	\end{aligned}
	\label{eq:dppa drives}
\end{equation}
that is, the first (second) cavity has two drives, close to the red (blue) motional sidebands corresponding to the mechanical resonators [cf.~\cref{fig:dppa}(a-b)].
We have already written the amplitudes in terms of cooperativities
$\cc_{1i}=4|G_{1i}|^2/(\kappa_1\Gamma_{\mathrm m,i})$, $\cc_{2i}=4|J_{2i}|^2/(\kappa_2\Gamma_{\mathrm m,i})$,
and chosen the cooperativities in both arms to be equal $\cc_1\equiv\cc_{1i}$, $\cc_2\equiv\cc_{2i}$.
Given \cref{eq:2by2,eq:dppa drives}, isolation ($\matr T_{12}=0$) requires $\delta_1^2\Gamma_{\mathrm m,2}^2=\delta_2^2\Gamma_{\mathrm m,1}^2$ \footnote{
	This condition causes the moduli of the transmission amplitudes via resonator 1 and 2 to coincide, $|G_{11}J_{21}\chi_{\mathrm m,1}(0)|=|G_{12}J_{22}\chi_{\mathrm m,2}(0)|$,
needed for complete destructive interference.}.
$\delta_1$ and $\delta_2$ must have opposite signs,
and we parametrize them by a single dimensionless variable $\delta_1=\delta\Gamma_{\mathrm m,1}$, $\delta_2=-\delta\Gamma_{\mathrm m,2}$.

Isolation occurs for certain phases of the coupling amplitudes 
$\theta_{1i}\equiv\arg(G_{1i})$ and $\theta_{2i}\equiv\arg(J_{2i})$.
However, only the overall relative ``plaquette phase'',
$\Phi\equiv\theta_{11}+\theta_{21}-\theta_{12}-\theta_{22}$, is relevant,
which explains the parameterization in \cref{eq:dppa drives}.
Setting $\delta=\sqrt{2\cc_1-1}/2$ achieves impedance matching
(i.e., vanishing reflection at cavity 1), attainable for $\cc_1\geq0.5$.
Then the plaquette phase at which isolation occurs is
\begin{equation}
  	\Phi=i\log\left( \frac{2\delta-i}{2\delta+i} \right)
    =2\arccos\sqrt{1-1/(2\cc_1)}.
  	\label{eq:plaquette phase}
\end{equation}
Inverting the plaquette phase $-\Phi$ leads to isolation in the opposite direction (cf.~SI).

We have chosen the couplings [\cref{eq:dppa drives}] due to the following reasons.
First, an even number of blue and red tones ensures equivalent arms of the circuit.
Second, amplification requires blue tones.
Third, a directional amplifier with four blue tones cannot be impedance matched to the signal source (cf.~SI).
Last, swapping hopping and amplifier interactions in one arm of the circuit cannot lead to directional amplification \footnote{Private communication with Anja Metelmann.
	Note that choosing $G_{21}=J_{22}=0$, $G_{11}=J_{12}$, $\delta=0$,
	yields reciprocal amplifier with unlimited gain-bandwidth product~\cite{Metelmann2014}.
}.

The condition $\cc_2<\cc_1$ ensures that the system does not exceed the parametric instability threshold.
In the limit of large gain, we obtain our first main result, the scattering matrix
\begin{equation}
  	\mat{a_{1,\text{out}}(0)\\a_{2,\text{out}}\dagg(0)}=
  	\mat{\frac{1}{\sqrt{2}}&-\frac{1}{\sqrt{2}}&0&0\\
  	\frac{i\sqrt{\G}}{\sqrt{4\cc_1}}&\frac{i\sqrt{\G}}{\sqrt{4\cc_1}}&-\sqrt{\G}&\frac{\cc_1+\cc_2}{\cc_2-\cc_1}}
  	\mat{b_{1,\text{in}}(0)\\b_{2,\text{in}}(0)\\a_{1,\text{in}}(0)\\a_{2,\text{in}}\dagg(0)},
	\label{eq:dppa S}
\end{equation}
with vanishing reverse gain $|S_{a_2\to a_1}(0)|^2$, but forward gain
\begin{equation}
    |S_{a_1\to a_2}(0)|^2\equiv\G=\frac{4\cc_1\cc_2}{(\cc_1-\cc_2)^2},
	\label{eq:dppa G}
\end{equation}
which can in principle be arbitrarily large, as long as the RWA is valid (cf.~SI).

At the same time, thermal noise from the mechanical resonators is suppressed by increasing $\cc_1$,
as is demonstrated in \cref{fig:dppa}(e), where we plot the noise added to the signal
$\N(\omega)=\G^{-1}\sum_{i\neq a_1}(n_{i}+1/2)|S_{i\to a_2}(\omega)|^2$ \cite{Caves1982,Clerk2010,Nunnenkamp2014},
where we sum over all noise sources, with associated thermal occupation $n_{i}$, and scattering amplitude to the second cavity $S_{i\to a_2}$.
Using \cref{eq:dppa S}, and denoting thermal cavity (mechanical) occupations by $n_{\mathrm c,i}$ ($n_{\mathrm m,i}$),
the noise on resonance
\begin{equation}
  	\N_{\text{DPPA}}=\frac{1}{4\cc_1}\left( n_{\mathrm m,1}+n_{\mathrm m,2}+1\right)+\frac{(\cc_1+\cc_2)^2}{4\cc_1\cc_2}\left( n_{\mathrm c,2}+\frac{1}{2} \right).
	\label{eq:dppa N}
\end{equation}
As a result, for large $\cc_1\gtrsim\cc_2$, and vanishing thermal occupation of the cavity input, we reach the quantum limit of half a quantum of added noise, $\N_{\text{DPPA}}\to1/2$~\cite{Caves1982,Clerk2010}.

Another important figure of merit is noise emerging from cavity 1, characterized by the output noise spectral density,
$S_{a_1}^{\text{out}}(\omega)\equiv\int\tfrac{\dd{\omega'}}{2\pi}\langle a_{1,\text{out}}\dagg(\omega) a_{1,\text{out}}(\omega')\rangle$, which we plot in \cref{fig:dppa}(f).
Ultimately, the reason for building directional amplifiers is to reduce this figure.
On resonance, $S_{1,\text{out}}^{\text{DPPA}}(0)=(n_{\mathrm m,1}+n_{\mathrm m,2}+1)/2$.
Strategies to reduce this figure are discussed below.

\begin{figure*}[t]
	\centering
	\includegraphics[width=\linewidth]{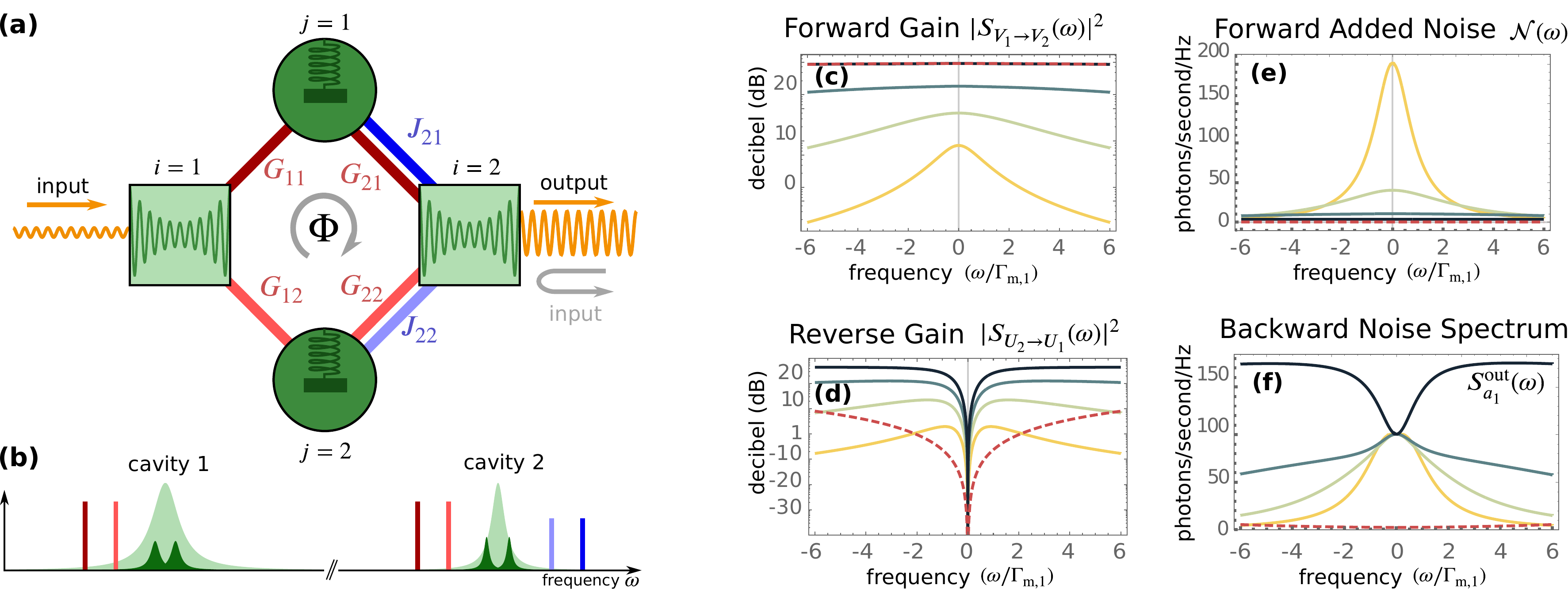}
	\caption{(Color online.) The directional phase-sensitive amplifier (DPSA).
		\textbf{(a)} Model. Single red hopping interactions are impedance matched, double red-blue provide the phase-sensitive amplification.
		\textbf{(b)} Pump setup. Cavity 1 is pumped on the red sidebands of the mechanical resonators, whereas cavity 2 has pumps on the red and blue sidebands.
		In \textbf{(c-f)} we plot forward gain, reverse gain [cf.~\cref{eq:dppa S}], added noise [\cref{eq:dppa N}], and the output noise fluctuation spectrum of cavity 1, all as functions of frequency in units of $\Gamma_{\mathrm m,1}$,
        for cooperativities $\cc_1=\{1,3,10,30\}$, $\cc_2=\cc_1^2$ (yellow to black, or light to dark).
		The parameters are the same as in \cref{fig:dppa}.
        Depending on parameters, external sideband cooling with an auxiliary mode can achieve $n_{\mathrm m,j}\approx0$, without negatively affecting amplification properties, as described below.
        The red (dashed) curve in each plot illustrates this case, with 
        $\cc_1=30$ and effective parameters $n_{\text{eff},i}=n_{\mathrm m,i}(\Gamma_{\mathrm m,i}/\Gamma_{\text{eff},i}),\Gamma_{\text{eff},i}=50\Gamma_{\mathrm m,i}$.
	}
	\label{fig:dpsa}
\end{figure*}

The off-resonance behavior of the DPPA is remarkably rich and depends on the dimensionless quantities $\kappa_i/\Gamma_{\mathrm m,j},\cc_1,\cc_2$.
We plot forward gain, reverse gain, added noise, and the noise spectrum at cavity 1 as functions of frequency at cooperativities $\cc_1=\{1,3,10,30\},\cc_2=\cc_1-0.1\sqrt{\cc_1}$ in \cref{fig:dppa}(c-f).
$\cc_2$ is chosen such that when increasing $\cc_1$ both gain and bandwidth are enhanced.

We show in the SI that for $\Gamma_{\mathrm m,j}=\Gamma_{\mathrm m}$ and $\kappa_i=\kappa$ and in the regime where $\kappa/\Gamma_{\mathrm m}\gg\{1,\cc_1,\cc_2\}$ the gain bandwidth is $\Gamma=4\sqrt{\cc_1\cc_2}\Gamma_{\mathrm m}$ [cf.~\cref{fig:dppa}(c)].
As the gain gets large and $\cc_1,\cc_2$ dominate all other dimensionless parameters, the bandwidth approaches $\Gamma=\kappa(\cc_1-\cc_2)/\cc_1$,
leading to the gain-bandwidth product limit $P\equiv\Gamma\sqrt{\G}\to2\kappa$,
independent of $\kappa/\Gamma_{\mathrm m}$.
Close to resonance, the reverse scattering amplitude $S_{a_2\to a_1}(\omega)\approx-i\omega\sqrt\G/\Gamma_{\mathrm m}$ (cf.~SI),
such that the product of isolation bandwidth and gain is $\Gamma_{\mathrm m}$.
Since the gain bandwidth is larger than the isolation bandwidth,
there is large reverse gain off resonance [cf.~\cref{fig:dppa}(d)], and noise from cavity 2 dominates the noise spectral density at cavity 1 [cf.~\cref{fig:dppa}(f)].
With increasing effective mechanical linewidth $\Gamma_{\mathrm m}$ (through additional sideband cooling),
the isolation bandwidth grows, suppressing reverse gain off resonance
(cf.\ red, dashed  curve in \cref{fig:dppa} and Ref.~\cite{Metelmann2015}).
In the SI we calculate how off-resonant terms renormalize the parameters of the DPPA.

\emph{Directional phase-sensitive amplifier (DPSA).}---We now turn to an implementation of a DPSA, which necessitates six tones.
Essentially, we replace the amplifier interaction in the DPPA by a phase-sensitive quantum non-demolition (QND) interaction that couples one quadrature of cavity 2 to only one quadrature of the mechanical resonator~\cite{Braginsky1980,Clerk2008}, choosing
\begin{equation}
  	\matr G=\frac12\mat{e^{i\Phi/2}\sqrt{\cc_1\Gamma_{\mathrm m,1}\kappa_1}&e^{-i\Phi/2}\sqrt{\cc_1\Gamma_{\mathrm m,2}\kappa_1}\\
    \sqrt{\cc_2\Gamma_{\mathrm m,1}\kappa_2}&\sqrt{\cc_2\Gamma_{\mathrm m,2}\kappa_2}},\\
	\label{eq:dpsa drives}
\end{equation}
and the same $\matr J$ as for the DPPA [\cref{eq:dppa drives}], illustrated in \cref{fig:dpsa}(a,b).
Since the QND interaction requires $|G_{i2}|=|J_{i2}|$, and we require symmetric amplifier arms,
two cooperativities suffice to characterize the six tones.
In the DPSA, we need to ensure that the measured mechanical quadratures agree, 
$\mu=\arg(G_{11}\chi_{\mathrm m,1}(0)G_{21}/J_{21})=\arg(\pm G_{12}\chi_{\mathrm m,2}(0)G_{22}/J_{22})$,
and that the information from both emerges in the same cavity quadrature, $\nu=\arg(G_{21}J_{21})=\arg(\pm G_{22}J_{22})$.
$\mu$ and $\nu$ determine the quadratures involved in amplification.
The two remaining phases are an arbitrary mechanical phase, and the plaquette phase.

While there is no parametric instability of the kind that limits back-action evading measurements~\cite{Hertzberg2010,Woolley2013},
we show in the SI using a Floquet technique \cite{Malz2016,Malz2016b} that counterrotating terms induce an instability threshold for finite sideband parameter (similar to Ref.~\cite{Li2015}), and the RWA is only valid for sideband parameters that are bigger than the cooperativities.
This is not out of reach~\cite{Teufel2011a}, but needs to be taken into account in experimental design.

The isolation, detuning, and impedance-matching conditions coincide with those of the DPPA, and we obtain another central result, the scattering matrix (on resonance)
\begin{multline}
  \mat{U_{1,\text{out}}\\V_{1,\text{out}}\\U_{2,\text{out}}\\V_{2,\text{out}}}
  =\mat{0&0&0&0\\0&0&0&0\\0&0&-1&0\\0&\sqrt\G&0&-1}
	\mat{U_{1,\text{in}}\\V_{1,\text{in}}\\U_{2,\text{in}}\\V_{2,\text{in}}}\\
	+\frac{1}{\sqrt{2}}\mat{1&0&-1&0\\0&1&0&-1\\0&0&0&0\\
	\sqrt{2\mathcal F}&0&\sqrt{2\mathcal F}&0}
  \mat{X_{1,\text{in}}\\P_{1,\text{in}}\\X_{2,\text{in}}\\P_{2,\text{in}}},
  \label{eq:dpsa S}
\end{multline}
where we defined noise scattering intensity $\mathcal F\equiv\frac{4\cc_2}{\cc_1^2}$, gain
\begin{equation}
  \G=\frac{8\cc_2(2\cc_1-1)}{\cc_1^2},
  \label{eq:dpsa G}
\end{equation}
mechanical $X_i=(b_i+b_i\dagg)/\sqrt{2}$, $P_i=i(b_i\dagg-b_i)/\sqrt{2}$, and
optical quadratures $U_i=(a_i+a_i\dagg)/\sqrt{2}$, $V_i=i(a_i\dagg-a_i)/\sqrt{2}$.

The amplifier is phase-sensitive and directional, as only the phase quadrature of the second cavity, $V_2$,
inherits the amplified signal from the phase quadrature of the first cavity, $V_1$.
We calculate the noise added to the signal as before
\begin{equation}
  \N_{\text{DPSA}}=\frac{n_{\mathrm m,1}+n_{\mathrm m,2}+1}{2(2\cc_1-1)}
  +\frac{\cc_1}{8\cc_2}\frac{\cc_1}{2\cc_1-1}\left( n_{\mathrm c,2}+\frac{1}{2} \right).
  \label{eq:dpsa N}
\end{equation}
The crucial difference to the DPPA is that the noise stemming from reflection of fluctuations at cavity 2 can \emph{also} be suppressed,
such that in the limit $\cc_2\gg\cc_1\gg1$ added noise vanishes.

To investigate the off-resonant behavior of the DPSA,
we plot forward gain, reverse gain, added noise, and spectral noise density at
cavity 1 in \cref{fig:dpsa}(c-f) at cooperativities $\cc_1=\{1,3,10,30\},\cc_2=\cc_1^2$.
Increasing $\cc_1$ enhances bandwidth and gain [cf.~\cref{fig:dpsa}(c)].
At the same time, the mechanical noise is suppressed [cf.~\cref{fig:dpsa}(e)].
Close to resonance, the reverse scattering behaves the same as for the DPPA
$S_{a_2\to a_1}(\omega)\approx-i\omega\sqrt\G/\Gamma_{\mathrm m}$ (cf.~SI),
and the same conclusions apply [cf.~\cref{fig:dpsa}(d,f)].
The gain-isolation-bandwidth product is $\Gamma_{\mathrm m}$.
Forward and reverse gain are proportional to $\sqrt{\cc_2}$, implying an unlimited gain-bandwidth product (cf.~SI). 
For equivalent mechanical resonators $\Gamma_{\mathrm m,1}=\Gamma_{\mathrm m,2}=\Gamma_{\mathrm m}$ and in the limit $\kappa/\Gamma_{\mathrm m}\gg\{1,\cc_1\}$,
the amplitude gain bandwidth of the DPSA is well approximated by $\Gamma_{\text{gain}}=2\cc_1\Gamma_{\mathrm m}$.

\emph{Backward propagating noise and sideband cooling.}---The noise emitted in the reverse direction is of central importance for directional amplifiers.
For both DPPA and DPSA, the output noise spectral density of cavity 1 on resonance is
$S_{1,\text{DPSA}}^{\text{out}}(0)=S_{1,\text{DPPA}}^{\text{out}}(0)
=(n_{\mathrm m,1}+n_{\mathrm m,2}+1)/2$.
Due to impedance matching and directionality fluctuations incident on the cavities do not appear in $a_{1,\text{out}}$ [cf.~\cref{eq:dppa S,eq:dpsa S}].
The commutation relations of $a_{1,\text{out}}$ then imply $\sum_{j=1}^2(|S_{b_i\to a_1}(0)|^2-|S_{b_i\dagg\to a_1}(0)|^2)=1$,
i.e,  mechanical fluctuations have to appear in the output instead.
The lowest possible value for $S_{1,\text{out}}$ is $1/2$, attainable for zero thermal noise quanta in the mechanical resonators.

However, even in state-of-the-art dilution refrigerators, the required temperatures are out of reach.
One way to mitigate backward noise emission is to add another microwave mode to the setup that can replace the fluctuations in the output of cavity 1, essentially realizing a circulator.
Without modifying the theory above, one can either increase the resonance frequency of the mechanical modes, which is mainly a technological challenge, or one could resort to external sideband cooling with an auxiliary mode.
The latter can achieve $n_{\mathrm m}\to0$~\cite{Teufel2011,Chan2011,Riviere2011},
and has the added benefit of enhancing mechanical linewidths [cf.\ red (dashed) curve in~\cref{fig:dppa,fig:dpsa} and SI].
Whilst this could be done with an additional cavity mode for each resonator,
implementing a circuit with four cavity modes coupled to two mechanical resonators is a formidable technical challenge.
A problem arises when cooling with only one additional mode,
since it can lead to a coupling of the mechanical resonators via the extra cooling mode,
thereby changing the topology of the system thus spoiling directionality.
This can be mitigated by detuning each pump by several mechanical linewidths (cf.~SI),
making cooling with only one additional mode feasible.

\emph{Conclusion.}---We have presented quantum-limited, nonreciprocal amplifiers using an optomechanical plaquette comprising two cavities and intermediate mechanical resonators~\cite{Bernier2017,Peterson2017}.
Such devices carry great promise, since they can be integrated into superconducting circuits
and amplify near or at the quantum limit, whilst protecting the signal source.

\begin{acknowledgments}
\emph{Acknowledgments.}---We are grateful to John Teufel, and in particular Anja Metelmann for insightful discussions and helpful comments.
DM acknowledges support by the UK Engineering and Physical Sciences Research Council (EPSRC) under Grant No. EP/M506485/1.
This work was supported by the SNF, the NCCR Quantum Science and Technology (QSIT), and the European Union's Horizon 2020 research and innovation programme under grant agreement No 732894 (FET Proactive HOT).
TJK acknowledges financial support from an ERC AdG (QuREM). AN holds a University Research Fellowship from the Royal Society and acknowledges support from the Winton Programme for the Physics of Sustainability.
\end{acknowledgments}

\bibliographystyle{apsrev4-1}
\bibliography{library}{}
  \begin{widetext}
\clearpage

\setcounter{equation}{0}
\setcounter{figure}{0}
\setcounter{table}{0}
\setcounter{page}{1}
\makeatletter
\renewcommand{\theequation}{S\arabic{equation}}
\renewcommand{\thefigure}{S\arabic{figure}}
\renewcommand{\bibnumfmt}[1]{[S#1]}
\renewcommand{\citenumfont}[1]{S#1}

\begin{center}
	\textbf{\large Supplemental Material: Quantum-limited directional amplifiers with optomechanics}
\end{center}
\section{Most general time-independent Hamiltonian}\label{app:Hamiltonian}
For an introduction to optomechanics and input-output theory, we refer to Refs.~\cite{Sgardiner2004quantum,SAspelmeyer2014}.
The optomechanical Hamiltonian for the plaquette is given by $(\hbar=1)$
\begin{equation}
	H_{\text{sys}}=\sum_i\omega_{\mathrm c,i}a_i\dagg a_i+\Omega_{i}b_i\dagg b_i-\sum_{ij}g_{0,ij}a_i\dagg a_i(b_j+b_j\dagg).
	\label{eq:bare general Hamiltonian}
\end{equation}
$g_{0,ij}$ is the bare coupling between the $i$th cavity and the $j$th mechanical resonator.
Each cavity is driven by up to four tones, at frequencies $\omega_{\mathrm c,i}\pm(\Omega_{j}+\delta_j)$, i.e.,
on the red and blue motional sidebands due to the mechanical resonators, as illustrated in Fig.~1 in the main text.
The drives generate coherent states in the cavities, such that the annihilation operators for the cavity modes can be written as a sum of coherent parts $\alpha_i$ and fluctuations $\delta \hat a_i$
\begin{equation}
	\hat a_i=e^{-i\omega_{\mathrm c,i} t}\left( \alpha_{i1+}e^{-i(\Omega_{1}+\delta_1)t}+\alpha_{i1-}e^{i(\Omega_{1}+\delta_1)t}
	+\alpha_{i2+}e^{-i(\Omega_{2}+\delta_2)t}+\alpha_{i2-}e^{i(\Omega_{2}+\delta_2)t}+\delta\hat  a_i\right)\equiv e^{-i\omega_{\mathrm c,i}t}[\alpha_i(t)+\delta \hat a_i],
	\label{eq:displacement}
\end{equation}
where $\alpha_i$ are c-numbers that describe the coherent state, and $\delta\hat  a_i$ is a bosonic operator with zero mean.
In the following, we will rename $\delta\hat  a_i\to\hat  a_i$ for a cleaner notation, and drop the hats.
Going into a rotating frame with respect to 
\begin{equation}
  H_0=\sum_j\left( \Omega_{j}+\delta_j \right)b_j\dagg b_j+\sum_i\omega_{\mathrm c,i}a_i\dagg a_i,
  \label{eq:H0}
\end{equation}
and using \cref{eq:displacement}, the time-independent Hamiltonian [Eq.~(1) in the main text] corresponds to all time-independent (resonant) terms in
\begin{equation}
  	H_{\text{rotating frame}}=-\sum_j\delta_jb_{j}\dagg b_j-\sum_{ij}g_{0,ij}\left\{\alpha_i^*(t) a_i\left[b_je^{-i(\Omega_{j}+\delta_j)t}+b_j\dagg e^{i(\Omega_{j}+\delta_j)t}\right]+\text{H.c.}\right\}.
  \label{eq:H rotating frame}
\end{equation}

In the main text, we neglect all time-dependent terms for simplicity.
When calibrating pumps and phases, their effect will have to be included.
Below, we answer three questions about counterrotating terms. First, do they change the topology of the circuit and therefore make directionality impossible? We find that to a good approximation this is not the case, as we analyze in section ``Off-resonant terms''.
Second, do off-resonant terms change the effective parameters of the circuit? As in previous studies the answer is yes, detailed in Sections ``Off-resonant terms'' and ``Stability''.
Third, can off-resonant terms lead to an instability? Again the answer is yes, for sufficiently strong driving or sufficiently low sideband parameter, as described in Section ``Stability''.
We also find that the RWA theory is valid for large but finite sideband parameter, after taking into account the modification of effective parameters.

We describe the system with input-output theory and quantum Langevin equations
\cite{SGardiner1985,Sgardiner2004quantum,SAspelmeyer2014}
\begin{subequations}
	\begin{align}
		\label{eq:mech eom}
		b_j(\omega)&=\chi_{\mathrm m,j}(\omega)\left\{i\sum_{i=1}^2\left[ a_i(\omega)G_{ij}^*+a_i\dagg(\omega)J_{ij} \right]+b_{j,\text{in}}(\omega)\right\},\\
		\label{eq:opt eom}
		a_i(\omega)&=\chi_{\mathrm c,i}(\omega)\left\{i\sum_{j=1}^2\left[ G_{ij}b_j(\omega)+J_{ij}b_j\dagg(\omega) \right]+a_{i,\text{in}}(\omega)\right\}
	\end{align}
	\label{eq:eom}
\end{subequations}
with susceptibilities $\chi_{\mathrm m,i}(\omega)=[\Gamma_{\mathrm m,i}/2-i(\omega+\delta_i)]^{-1}$ and $\chi_{\mathrm c,i}(\omega)=[\kappa_i/2-i\omega]^{-1}$, 
and mechanical (cavity) dissipation rates $\Gamma_{\mathrm m,i}$ ($\kappa_i$).
The mechanical (cavity) input noise operators $b_{i,\text{in}}$ ($a_{i,\text{in}}$)
are assumed to have bosonic commutation relations and delta-correlated noise
$\langle b_{i,\text{in}}\dagg(t) b_{j,\text{in}}(t')\rangle=\delta_{ij}n_{\mathrm m,i}\delta(t-t')$, $\langle a_{i,\text{in}}\dagg(t) a_{j,\text{in}}(t')\rangle=\delta_{ij}n_{\mathrm c,i}\delta(t-t')$.
In order to obtain a description only in terms of the microwave modes, we eliminate the mechanical degrees of freedom in \cref{eq:opt eom} with \cref{eq:mech eom}.
This yields Eq.~(3) in the main text.

The coupling matrices $i\matr T_{ij}$, defined in Eq.~3, exhibit the symmetry
\begin{equation}
	\matr T_{ij}(\omega)=\matr\sigma_1\matr T_{ij}^*(-\omega)\matr\sigma_1,
	\label{eq:symmetry}
\end{equation}
since we are using both annihilation operators and their conjugates.
The minus sign in the frequency is due to our choice of Fourier transform, $[a_i(\omega)]\dagg=a_i\dagg(-\omega)$.
This is the only symmetry, since we know that there are 8 free complex parameters (the 8 driving amplitudes),
and $T$ has $4^2=16$ complex entries.

\section{More details on detuning choices}
Isolation can be achieved when $\matr T_{12}=0$, i.e., the first cavity is decoupled from the second, 
but $\matr T_{21}\neq0$, i.e., the second cavity \emph{is} coupled to the first.
For the DPPA, these requirements turn into
\begin{subequations}
	\begin{align}
      [i\matr T_{12}]_{12}&=\chi_{\mathrm m,1}(0)G_{11}J_{12}+\chi_{\mathrm m,2}(0)G_{21}J_{22}=0,\\
      [i\matr T_{21}]_{12}&=\chi_{\mathrm m,1}^*(0)G_{11}J_{12}+\chi_{\mathrm m,2}^*(0)G_{21}J_{22}\neq0.
	\end{align}
    \label{eq:isolation condition}
\end{subequations}
For the choice of driving strengths in the main text Eq.~(5) (repeated here for convenience)

\begin{equation}
  \matr G=\frac12\mat{e^{\frac{i\Phi}{2}}\sqrt{\cc_1\Gamma_{\mathrm m,1}\kappa_1}&0\\
  e^{-\frac{i\Phi}{2}}\sqrt{\cc_1\Gamma_{\mathrm m,2}\kappa_1}&0},\quad
  \matr J=\frac12\mat{0&\sqrt{\cc_2\Gamma_{\mathrm m,1}\kappa_2}\\0&\sqrt{\cc_2\Gamma_{\mathrm m,2}\kappa_2}},
\end{equation}
we find
\begin{subequations}
\begin{align}
    0&=i\Gamma_{\mathrm m,1}\Gamma_{\mathrm m,2}+2\delta_1\Gamma_{\mathrm m,2}+e^{i\Phi}(i\Gamma_{\mathrm m,1}\Gamma_{\mathrm m,2}+2\delta_2\Gamma_{\mathrm m,1}),\\
    0&\neq i\Gamma_{\mathrm m,1}\Gamma_{\mathrm m,2}+2\delta_1\Gamma_{\mathrm m,2}+e^{-i\Phi}(i\Gamma_{\mathrm m,1}\Gamma_{\mathrm m,2}+2\delta_2\Gamma_{\mathrm m,1}).
\end{align}
  \label{eq:isolation conditions2}
\end{subequations}
Taking the modulus, we find that isolation requires
\begin{equation}
  \frac{\delta_1^2}{\Gamma_{\mathrm m,1}^2}=\frac{\delta_2^2}{\Gamma_{\mathrm m,2}^2}.
\end{equation}
Note that sending $\Phi\to-\Phi$ interchanges the two equations in \cref{eq:isolation conditions2}, which means that isolation now occurs in the opposite direction.
In order to avoid ``double isolation'', where the forward and backward transmission vanish at the same plaquette phase
(as per the previous sentence, this occurs for $\Phi=0$ or $\Phi=\pi$),
we need $\delta_1/\Gamma_{\mathrm m,1}\neq\delta_2/\Gamma_{\mathrm m,2}$.
The phase at which isolation is obtained is given in Eq.~(6) in the main text.

For these choices, the scattering amplitude from $a_{1,\text{in}}$ to $a_{1,\text{out}}$ (reflection) is given by
\begin{equation}
  S_{a_1\to a_1}(0)=\frac{4\cc_1}{4\delta^2+2\cc_1+1}-1.
  \label{eq:reflection}
\end{equation}
Impedance matching is attained at zero reflection, namely when 
\begin{equation}
  \delta=\sqrt{2\cc_1-1}/2,
  \label{eq:detuning choice}
\end{equation}
the same as in Ref.~\cite{SPeterson2017}.

\section{Directional phase-preserving amplifier with only blue tones}
Here we analyze the optomechanical plaquette with only pumps on the upper motional sidebands. 
While directional phase-preserving amplification is still possible, the signal cannot be impedance matched.
We choose coupling amplitudes as follows
\begin{equation}
	\matr G=0,\qquad
    \matr J=\frac12\mat{e^{\frac{i\Phi}{2}}\sqrt{\cc_1\Gamma_{\mathrm m,1}\kappa_1}&e^{-\frac{i\Phi}{2}}\sqrt{\cc_1\Gamma_{\mathrm m,2}\kappa_1}\\
    \sqrt{\cc_2\Gamma_{\mathrm m,1}\kappa_2}&\sqrt{\cc_2\Gamma_{\mathrm m,2}\kappa_2}}.
\end{equation}
The analysis proceeds similarly to the DPPA in the main text. $\Phi$ is the only physically relevant phase. 
We choose the same detuning parameterization as before $\delta_1=\delta\Gamma_{\mathrm m,1}$ and $\delta_2=-\delta\Gamma_{\mathrm m,2}$, in which case the plaquette phase takes the same form as before [Eq.~(6)].
However, if we look at the optical scattering matrix, we find
\begin{equation}
	\mat{a_{1,\text{out}}\\a_{2,\text{out}}}
    =\mat{-1-\frac{4\cc_1}{4\delta^2+1-2\cc_1}&0\\
    \frac{-16\delta\sqrt{\cc_1\cc_2(4\delta^2+1)}}{(4\delta^2-2\cc_1+1)(4\delta^2-2\cc_2+1)}&-1-\frac{4\cc_2}{4\delta^2-2\cc_2+1}}
	\mat{a_{1,\text{in}}\\a_{2,\text{in}}} + \text{mechanical noise}.
\end{equation}
The cooperativities are obey $4\delta^2+1>\cc_i>0$, where the first condition is required for stability and the second arises by definition, such that impedance matching is not possible, and input will be reflected and amplified.
This property is already highly undesirable in a directional amplifier, since the main point of a directional amplifier is protection of the system that emits the signal.

Amplification is obtained if either or both of the cooperativities approach $2\delta^2+1/2$.
$\cc_1\to2\delta^2+1/2$ leads to a lot of noise being emitted from cavity 1, both due to the reflection and due to amplified mechanical noise (not shown above).
Hence, we consider $\cc_2=2\delta^2+1/2-\eps$, for some small $\eps>0$, and $\cc_1=1/2$.
For $\delta^2\gg\eps$, we can then write 
\begin{equation}
	\mat{a_{1,\text{out}}\\a_{2,\text{out}}}
    =\frac{\sqrt{4\delta^2+1}}{\sqrt{2}\delta}\mat{-\frac{1}{2\delta}&\frac{1}{2\delta}\\
    -\frac{\delta(2\delta+i)+1}{\eps}&\frac{\delta(2\delta-i)+1}{\eps}}\mat{b_{1,\text{in}}\\b_{2,\text{in}}}
    -\mat{1+\frac{1}{2\delta^2}&0\\
    \frac{4\delta^2+1}{\delta\eps}&\frac{4\delta^2+1}{\eps}}
	\mat{a_{1,\text{in}}\\a_{2,\text{in}}}.
\end{equation}
For large $\delta$, the signal source is only subject to the reflected signal and noise at cavity 1,
leading to the approximate scattering matrix
\begin{equation}
	\mat{a_{1,\text{out}}\\a_{2,\text{out}}}
    =\frac{\delta^2}{\eps}\mat{0&0\\-1&1}\mat{b_{1,\text{in}}\\b_{2,\text{in}}}
    +\mat{-1&0\\\frac{4\delta}{\eps}&\frac{4\delta^2}{\eps}}
	\mat{a_{1,\text{in}}\\a_{2,\text{in}}}.
\end{equation}
The suppression of mechanical noise works well for cavity 1, but cavity 2, where the signal emerges, is subject to amplified mechanical noise.
As a consequence, the amplification is not quantum-limited, in addition to the lack of impedance matching.
Interchanging $\cc_1\leftrightarrow\cc_2$ leads to
\begin{equation}
	\mat{a_{1,\text{out}}\\a_{2,\text{out}}}
    =\frac{4\delta}{\sqrt2\eps}\mat{-1&1\\-1&1}\mat{b_{1,\text{in}}\\b_{2,\text{in}}}
    +\mat{-\frac{4\delta^2}{\eps}&0\\-\frac{4\delta}{\eps}&-1}
	\mat{a_{1,\text{in}}\\a_{2,\text{in}}},
\end{equation}
i.e., the amplification is quantum-limited, but the signal source is subject to amplified mechanical and optical noise.

\section{Bandwidth and gain-bandwidth product of DPPA}
For $\Gamma_{\mathrm m,1}=\Gamma_{\mathrm m,2}=\Gamma_{\mathrm m}$ and $\kappa_1=\kappa_2=\kappa$, the scattering amplitudes are
\begin{subequations}
  \begin{align}
	S_{a_2\dagg\to a_1}(\omega)&=(-i\omega)\sqrt{2\cc_2(2\cc_1-1)}/\mathcal A(\omega),\\
	S_{a_1\to a_2\dagg}(\omega)&=\sqrt{2\cc_2(2\cc_1-1)}(i\omega-\Gamma_{\mathrm m})/\mathcal A(\omega),
  \end{align}
  \label{eq:dppa scattering amplitudes}
\end{subequations}
with
\begin{equation}
  \mathcal A(\omega)=(\Gamma_{\mathrm m}\kappa^2)^{-1}\{i\omega(\Gamma_{\mathrm m}-i\omega)(\kappa-2i\omega)^2+\cc_1\Gamma_{\mathrm m}(\kappa-2i\omega)[\Gamma_{\mathrm m}\kappa-i(\Gamma_{\mathrm m}+\kappa)\omega]+\cc_2\Gamma_{\mathrm m}\kappa[2\omega^2+i(\Gamma_{\mathrm m}+\kappa)\omega-\Gamma_{\mathrm m}\kappa]\}.
  \label{eq:AA}
\end{equation}
For the gain-bandwidth product, we are most interested in the limit of large gain, $\G\gg1$.
This implies $(\cc_1-\cc_2)^2\ll4\cc_1\cc_2<2(\cc_1+\cc_2)^2$.
We further assume $\cc_1\gg\{1,\Gamma_{\mathrm m}/\kappa,\kappa/\Gamma_{\mathrm m}\}$, in which case the first term in the curly brackets in \cref{eq:AA} can be neglected.
In this approximation,
\begin{equation}
	\mathcal A(\omega)=\kappa^{-2}\left\{ (\cc_1+\cc_2)\left[ -i\omega\Gamma_{\mathrm m}(\kappa-i\omega) \right]
    +(\cc_1-\cc_2)\left[\Gamma_{\mathrm m}(\kappa-i\omega)^2-i\kappa\omega(\kappa-2i\omega)\right]\right\}.
\end{equation}
The bandwidth is approximated by the smallest $|\omega|$ at which $2|\mathcal A(0)|^2=|\mathcal A(\omega)|^2$.
Expanding both square brackets to first order in $\omega$, we find that the amplitude bandwidth is approximately
\begin{equation}
	\Gamma=\frac{2\Gamma_{\mathrm m}(\cc_1-\cc_2)}{2\cc_1\frac{\Gamma_{\mathrm m}}{\kappa}+\left(1+\frac{\Gamma_{\mathrm m}}{\kappa}\right)(\cc_1-\cc_2)}.
\end{equation}
The gain-bandwidth product $P\equiv \Gamma\sqrt{\G}$ is
\begin{equation}
	P=\frac{2\Gamma_{\mathrm m}\sqrt{4\cc_1\cc_2}}{2\cc_1\frac{\Gamma_{\mathrm m}}{\kappa}+\left(1+\frac{\Gamma_{\mathrm m}}{\kappa}\right)(\cc_1-\cc_2)}.
\end{equation}
In the limit of large gain, with $\cc_2\to\cc_1$ and $2\cc_1\gg\cc_1-\cc_2$, $P$ tends to $2\kappa$.

We will now analyze the limits $\kappa\gg\Gamma_{\mathrm m}$ and $\kappa\ll\Gamma_{\mathrm m}$,
but note that they are only valid as long as $\cc_1$
is smaller than the ratio of $\kappa/\Gamma_{\mathrm m}$ and $\Gamma_{\mathrm m}/\kappa$.
In the limit $\kappa/\Gamma_{\mathrm m}\gg\{1,\cc_1,\cc_2\}$, we obtain
\begin{equation}
	\mathcal A(\omega)\approx (\Gamma_{\mathrm m}-i\omega)\Gamma_{\mathrm m}^{-1}[\Gamma_{\mathrm m}(\cc_1-\cc_2)+i\omega],
\end{equation}
which yields an amplitude bandwidth
\begin{equation}
	\Gamma=2(\cc_1-\cc_2)\Gamma_{\mathrm m}
    \label{eq:dppa bandwidth}
\end{equation}
and intensity bandwidth $\Gamma/2$.
The gain-bandwidth product, defined as $P\equiv\Gamma\sqrt{\G}$ evaluates to
\begin{equation}
	P=2\sqrt{4\cc_1\cc_2}\Gamma_{\mathrm m}.
\end{equation}
This implies that the gain-bandwidth product $P$ increases with gain.
When $\cc_1\Gamma_{\mathrm m}\sim\kappa$ the approximation above breaks down.

In the opposite limit, $\kappa\ll\Gamma_{\mathrm m}$, we instead have
\begin{equation}
	\mathcal A(\omega)\approx\kappa^{-2}\Gamma_{\mathrm m}(\kappa-i\omega)\cc_1^2[\kappa(\cc_1-\cc_2)/(2\cc_1)-i\omega],
\end{equation}
which implies that the bandwidth is close to $\kappa(\cc_1-\cc_2)/\cc_1$ and thus $P\approx \kappa\sqrt{4\cc_2/\cc_1}$.
For large gain, $P\approx 2\kappa$, as before.

The isolation bandwidth must be calculated separately. It is the range of frequencies over which sufficient isolation is attained.
What sufficient means has to decided depending on the purpose.
Close to $\omega=0$, to lowest order in $\omega$, the reverse gain departs linearly from zero, namely
\begin{equation}
	S_{a_2\dagg\to a_1}(\omega)
    =-i\omega\frac{\sqrt{2\cc_2(2\cc_1-1)}}{\Gamma_{\mathrm m}(\cc_1-\cc_2)}+\O(\omega^2)=-i\omega\sqrt{\G}/\Gamma_{\mathrm m}+\O(\omega^2).
\end{equation}
Thus, the isolation bandwidth is of order $\Gamma_{\mathrm m}/\sqrt{\G}$, independent of $\kappa$.

\section{Bandwidth and gain-bandwidth product of DPSA}
\label{app:dpsa bandwidth}
For the DPSA as discussed in the main text, we obtain
\begin{subequations}
  \begin{align}
  	S_{a_1\to a_2\dagg}(\omega)&=\sqrt{2\cc_2(2\cc_1-1)}(i\omega-\Gamma_{\mathrm m})/\mathcal B(\omega),\\
  	S_{a_2\dagg\to a_1}(\omega)&=\sqrt{2\cc_2(2\cc_1-1)}(-i\omega)/\mathcal B(\omega),\\
    \text{with}\qquad  	\mathcal B(\omega)&=(\Gamma_{\mathrm m}\kappa^2)^{-1}(\kappa-i\omega)\{
    \cc_1\Gamma_{\mathrm m}[\Gamma_{\mathrm m}\kappa-i\omega(\Gamma_{\mathrm m}+\kappa)]
    -(\kappa-2i\omega)i\omega(\Gamma_{\mathrm m}-i\omega)\}.
  \end{align}
\end{subequations}
Here, since both $S_{2\to1}\propto\sqrt{\cc_2}$ and $S_{1\to2}\propto\sqrt{\cc_2}$,
we can immediately conclude that the bandwidth is independent of the gain, and the gain-bandwidth product therefore unlimited.

For $\{\cc_1\omega,\cc_1\Gamma_{\mathrm m}\}\ll\kappa$, we find
\begin{equation}
	\mathcal B(\omega)\approx(\Gamma_{\mathrm m}-i\omega)\Gamma_{\mathrm m}^{-1}(\cc_1\Gamma_{\mathrm m}-i\omega),
\end{equation}
and therefore
\begin{equation}
    S_{1\to2}(\omega)\approx\frac{\sqrt{2\cc_2(2\cc_1-1)}\Gamma_{\mathrm m}}{-(\cc_1\Gamma_{\mathrm m}-i\omega)},\qquad
  	S_{2\to1}(\omega)\approx\frac{\sqrt{2\cc_2(2\cc_1-1)}i\omega\Gamma_{\mathrm m}}{-(\Gamma_{\mathrm m}-i\omega)(\cc_1\Gamma_{\mathrm m}-i\omega)},
\end{equation}
such that the gain bandwidth $\Gamma_{1\to2}=2\cc_1\Gamma_{\mathrm m}$.

To gain information about the departure from isolation, we expand the reverse gain around $\omega=0$. 
To first order (note that we do \emph{not} assume $\{\cc_1\omega,\cc_1\Gamma_{\mathrm m}\}\ll\kappa$ here)
\begin{equation}
	S_{2\to1}(\omega)=
    -i\omega\frac{\sqrt{2\cc_2(2\cc_1-1)}}{\cc_1\Gamma_{\mathrm m}}+\O(\omega^2)=-i\omega\sqrt{\G}/\Gamma_{\mathrm m}+\O(\omega^2).
\end{equation}
Thus, the isolation bandwidth is again of order $\Gamma_{\mathrm m}/\sqrt{\G}$ (but note that $\G$ takes different forms for DPSA and DPPA),
independent of $\kappa$.

\section{Off-resonant terms}
There are two kinds of off-resonant terms contained in \cref{eq:H rotating frame}.
One are ``counterrotating terms'', with a time-dependence $\mathcal O(2\Omega_{i})$, which are usually negligible. 
The other are off-resonant terms rotating at frequency $\Omega=\Omega_{2}+\delta_2-\Omega_{1}-\delta_1$, which
can have an appreciable effect~\cite{SBernier2017,SPeterson2017}.
For a related study see Ref.~\cite{SWoolley2013}.
The Hamiltonian describing the off-resonant terms is
\begin{equation}
  \begin{aligned}
  	H_{\text{off-resonant}}(t)&=-\sum_ia_i\dagg
  	\left[ g_{0,i2}\alpha_{i1+}b_2\dagg e^{i\Omega t}+g_{0,i2}\alpha_{i1+}b_2e^{-i\Omega t}+g_{0,i1}\alpha_{i2+}b_1\dagg e^{-i\Omega t}+g_{0,i1}\alpha_{i2+}b_1e^{i\Omega t} \right]+\text{H.c.}\\
  	&=-\sum_ia_i\dagg\left[ e^{i\Omega t}\left( \tilde J_{i2}b_2\dagg+\tilde G_{i1}b_1 \right)
  	+e^{-i\Omega t}\left( \tilde G_{i2}b_2+\tilde J_{i1}b_1\dagg \right)\right]+\text{H.c},
  \end{aligned}
  \label{eq:off-resonant}
\end{equation}
where $\tilde G_{i1}=G_{i2}g_{0,i1}/g_{0,i2}$ (and the same for $G\leftrightarrow J$ and $1\leftrightarrow2$).
Including $H_{\text{off-resonant}}$, the Hamiltonian is no longer time-independent, but rather periodic, with period $2\pi/\Omega$. 
The resulting explicitly time-dependent Langevin equations can be mapped to stationary ones by use of a Floquet formalism~\cite{SMalz2016,SMalz2016b},
where we write system operators as Fourier series, for instance $b(t)=\sum_n\exp(in\Omega t)b^{(n)}(t)$.
We obtain Langevin equations without explicit time-dependence (diagonal in frequency space)
\begin{subequations}
  \begin{align}
  	\chi_{\mathrm m,j}^{-1}(\omega-n\Omega)b_j^{(n)}(\omega)&=
  	i\sum_{i=1}^2\left( a_i^{(n)}G_{ij}^*+a_i^{(n)\dag}J_{ij} \right)+\delta_{n,0}\sqrt{\Gamma_{\mathrm m,j}}b_{j,\text{in}}
  	+i\sum_{i=1}^2\left( a_i^{(n\pm1)}\tilde G_{ij}^*+a_i^{(n\pm1)\dag}\tilde J_{ij} \right),\\
  	\chi_{\mathrm c,i}^{-1}(\omega-n\Omega)a_i^{(n)}(\omega)&=
  	i\sum_{j=1}^2\left( G_{ij}b_j^{(n)}+J_{ij}b_j^{(n)\dag} \right)+\delta_{n,0}\sqrt{\kappa_i}a_{i,\text{in}}
  	+i\left( \tilde J_{i2}b_2^{(n-1)\dag}+\tilde G_{i1}b_1^{(n-1)}+\tilde G_{i2}b_2^{(n+1)}+\tilde J_{i1}b_1^{(n+1)\dag} \right),
  \end{align}
  \label{eq:Floquet eoms}
\end{subequations}
where in \cref{eq:Floquet eoms}(a), $j=1$ corresponds to the $+$-sign and $j=2$ to the $-$-sign.
Note that in our conventions $[b^{(n)}(\omega)]\dagg=b^{(-n)}(-\omega)$.
In principle, Eqs.~\eqref{eq:Floquet eoms} constitute an infinitely large set of coupled linear equations (equivalently, an infinitely large matrix to invert). 
However, the coupling between different Fourier modes is suppressed by the mechanical and optical susceptibilities.
In particular, the mechanical susceptibilities are strongly peaked ($\Gamma_{\mathrm m,i}\ll\Omega$), such that it is a good approximation to let $b_j^{(n\neq0)}=0$.
Since in this approximation $b_j=b_j^{(0)}$, we will omit the superscript $(0)$ for $b$ in the following.
Another consequence of the approximation is $a_i^{(n)}=0$ for $|n|>1$.

We can think of $a_i^{(\pm1)}$ as four extra, ``virtual'' modes, as is done in Ref.~\cite{SPeterson2017}.
They are given by
\begin{subequations}
	\begin{align}
		a_i^{(1)}(\omega)&=i\chi_{\mathrm c,i}(\omega-\Omega)\left[\tilde J_{i2}b_2\dagg(\omega)+\tilde G_{i1}b_1(\omega)\right],\\
		a_i^{(-1)}(\omega)&=i\chi_{\mathrm c,i}(\omega+\Omega)\left[\tilde J_{i1}b_1\dagg(\omega)+\tilde G_{i2}b_2(\omega)\right],
	\end{align}
\end{subequations}
whence we write down the new equation of motion for $b$ in Fourier space, e.g.,
\begin{equation}
	\chi_{\mathrm m,1}^{-1}(\omega)b_1(\omega)=i\sum_i\left[G_{i1}^*a_i^{(0)}+J_{i1}a_i^{(0)\dag}\right]
    +\sqrt{\Gamma_{\mathrm m,1}}b_{1,\text{in}}
    -\sum_i\chi_{\mathrm c,i}(\omega-\Omega)\left[(|\tilde G_{i1}|^2-|\tilde J_{i1}|^2)b_1
    +(\tilde G_{i1}^*\tilde J_{i2}-\tilde J_{i1}\tilde G_{i2}^*)b_2\dagg\right].
\end{equation}
The two types of terms that appear due to the off-resonant terms is one proportional to $b_1$ that describes off-resonant cooling or heating,
which can be incorporated into the susceptibility of the mechanical resonator,
but also one that couples the first mechanical resonator to the second.
The latter process only occurs when there is a drive on the red sideband of one resonator and one on the blue sideband of the other resonator.
This is most easily understood when looking for example at the process underlying the term $\chi_{\mathrm m,1}(\omega-\Omega)b_2\dagg\tilde G_{11}^*\tilde J_{12}$.
$\tilde J_{12}$ is an interaction that creates a phonon in resonator 2 and a photon in cavity 1, but the process is off-resonant, meaning that the frequency of the photon created is approximately $\omega_{\mathrm c,1}-\Omega$.
$\tilde G_{11}$ shifts the frequencies the other way, such that this term mediates a resonant interaction between the mechanical resonators, with an off-resonant intermediate state.
In contrast, $\tilde J_{11}$ would create a phonon in resonator 1 and a photon in cavity at frequency $\omega_{\mathrm c,1}+\Omega$.
Thus the term $\tilde J_{11}\tilde J_{12}$ would produce a phonon with frequency $\Omega_{1}-2\Omega$,  a process that is strongly suppressed.

The spurious coupling between the resonators trivially vanishes in four-tone schemes, where one of $G_{ij}$ or $J_{ij}$ is zero for all $i,j$.
For more tones, we have to find $\tilde G_{i1}^*\tilde J_{i2}-\tilde J_{i1}\tilde G_{i2}^*=G_{i1}^*J_{i2}-J_{i1}G_{i2}^*$.
This vanishes for the DPSA, since in the case $i=1$, there are only red drives, and for $i=2$, the coupling strengths are the same $J_{2j}=G_{2j}$ [see Eq.~(10) in the main text].
Thus we can eliminate the off-resonant Fourier modes and write their effect as a self energy that modifies the susceptibility $\chi_{\mathrm m,i}^{-1}(\omega)\to\chi_{\mathrm m,i}^{-1}(\omega)+\Sigma_1(\omega)$, with
\begin{equation}
	\Sigma_1(\omega)=\sum_i\left(\frac{g_{0,i1}}{g_{0,i2}}\right)^2\chi_{\mathrm c,i}(\omega-\Omega)
    \left(|G_{i2}|^2-|J_{i2}|^2\right)
\end{equation}
(for $b_2$, we do $1\leftrightarrow2$).
For the DPPA, we obtain
\begin{equation}
  \begin{aligned}
      \Sigma_1^{\text{DPPA}}&=\left(\frac{g_{0,11}}{g_{0,12}}\right)^2\chi_{\mathrm c,1}(\omega-\Omega)\frac{\cc_1\Gamma_{\mathrm m,2}\kappa_1}{4}
      -\left(\frac{g_{0,21}}{g_{0,22}}\right)^2\chi_{\mathrm c,2}(\omega-\Omega)\frac{\cc_2\Gamma_{\mathrm m,2}\kappa_2}{4},\\
      \Sigma_2^{\text{DPPA}}&=\left(\frac{g_{0,12}}{g_{0,11}}\right)^2\chi_{\mathrm c,1}(\omega-\Omega)\frac{\cc_1\Gamma_{\mathrm m,1}\kappa_1}{4}
      -\left(\frac{g_{0,22}}{g_{0,21}}\right)^2\chi_{\mathrm c,2}(\omega-\Omega)\frac{\cc_2\Gamma_{\mathrm m,1}\kappa_2}{4},
  \end{aligned}
\end{equation}
whereas for the DPSA, the self-energies read
\begin{equation}
  \begin{aligned}
      \Sigma_1^{\text{DPSA}}&=\left(\frac{g_{0,11}}{g_{0,12}}\right)^2\chi_{\mathrm c,1}(\omega-\Omega)\frac{\cc_1\Gamma_{\mathrm m,2}\kappa_1}{4},\\
      \Sigma_2^{\text{DPSA}}&=\left(\frac{g_{0,12}}{g_{0,11}}\right)^2\chi_{\mathrm c,1}(\omega-\Omega)\frac{\cc_1\Gamma_{\mathrm m,1}\kappa_1}{4}.
  \end{aligned}
  \label{eq:dpsa self}
\end{equation}
Eqs.~\eqref{eq:dpsa self} do not have a contribution from the drives on the second cavity,
because the blue and red drives are balanced, such that the dynamical backaction cancels.

In the end, in an approximation where we neglect the frequency dependence of the self energies [i.e., $\chi_{\mathrm c,i}\approx (\kappa_i/2+i\Omega)^{-1}$],
which is valid for small frequencies around resonance $\omega\ll\Omega,\kappa_i$,
the effect of the complex self-energies can be subsumed as a change of damping and detuning parameters.

We stress again that the other important conclusion from this analysis is that the off-resonant terms do not change the topology of the circuit, neither in the DPPA nor in the DPSA,
such that the theory presented in the main text applies to a very good approximation.
The embedding into a Floquet ansatz explains how the extra modes and their properties arise in the description of Ref.~\cite{SPeterson2017}.

\section{Off-resonant cooling with auxiliary mode}
\begin{figure}[ht]
	\centering
	\includegraphics[width=\textwidth]{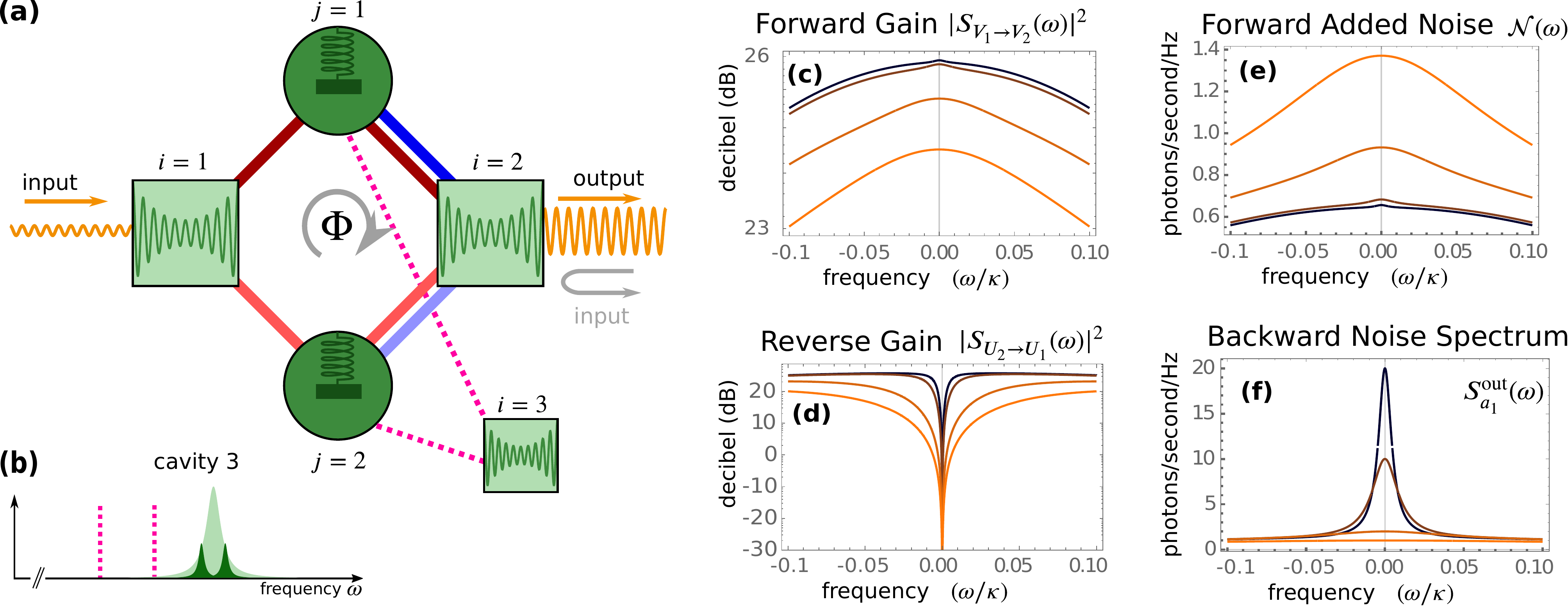}
	\caption{Off-resonant cooling with one extra mode.
	(a) The plaquette is extended with another cavity mode. It is clear from this figure that coupling this cavity mode to both mechanical resonators changes the topology of the plaquette.
	This is why we go into some length to show that detuning the extra mode by a sufficient amount mitigates this problem.
	(b) The third cavity mode is pumped with two tones close to the red sidebands of mechanical oscillators,
	at frequencies $\omega_{\mathrm c,3}-\Omega_j+\Delta_j$, with detunings $\Delta_1=-\Delta_2=\Delta\gg\Gamma_{\mathrm m,j}$.
	The detuning ensures that phonons are prevented from hopping from one resonator to the other.
	(c-f) We plot forward gain, reverse gain, forward added noise, and backward noise spectrum for \emph{fixed coupling strengths} (i.e., not fixed cooperativities), but with increasingly large off-resonant cooling.
	Parameters are $\kappa_2=\kappa_1=\kappa$, $\Gamma_{\mathrm m,1}=\Gamma_{\mathrm m,2}=\Gamma=10^{-4}\kappa\times\Lambda$, $\delta_1=\gamma_1 \delta,\delta_2=-\gamma_2\delta,\delta=\sqrt{2\cc_1-1},$ $\cc_1=4G_{11}^2/(\kappa\gamma)$, $G_{11}=\exp(i\Phi/2)0.2\kappa,G_{21}\exp(-i\Phi/2)0.2\kappa$, $G_{12}=G_{22}=J_{12}=J_{22}=2\kappa, J_{11}=J_{21}=0$, $n_{\mathrm c,1}=n_{\mathrm c_2}=0$, $n_{\mathrm m,1}=n_{\mathrm m,2}=1000/\Lambda$.
	Cooling parameter takes the values $\Lambda=\{50,100,500,1000\}$ as the color of the curve varies from black to orange.
	Note that we plot these curves as functions of frequency in units of $\kappa$, unlike the main text.
	}
	\label{fig:cooling}
\end{figure}
The above calculation can be repeated for the proposed off-resonant cooling with a third cavity mode. 
This involves two drives detuned from the sidebands of the mechanical resonator (or one drive roughly in the middle of the two sidebands).
For instance, consider pumps at frequencies $\omega_{\mathrm c,3}-\Omega_j+\Delta_j$, where $\omega_{\mathrm c,3}$ is the frequency of the third cavity mode, $\Delta_1=\Delta$, $\Delta_2=-\Delta$, and $\Delta\gg\Gamma_{\mathrm m,j}$.
This leads to similar off-resonant terms as above, namely,
\begin{equation}
	H_{\text{cool}}=-a_3\dagg\left( e^{-i\Delta t}G_{31}b_1 + e^{i\Delta t}G_{32}b_2\right)+\text{H.c.}
\end{equation}
This adds another contribution to the self-energy, similar to above
\begin{equation}
	\Sigma^{\text{cool}}_j(\omega)=\chi_{\mathrm c,3}(\omega)|G_{3j}|^2\approx \frac{2}{\kappa_3}|G_{3j}|^2.
\end{equation}
The detuning choice ensures that phonons cannot hop from one resonator to the other via the auxiliary mode.
The final result of this treatment is a new, enhanced damping rate $\Gamma_{\text{eff}}$. 
However, the noise strength in the Langevin equations is unchanged. 
We can write
\begin{equation}
	\dot b=\ldots+\sqrt{\Gamma_{\mathrm m}}b_{\text{in}}=\ldots+\sqrt{\Gamma_{\text{eff}}}\tilde b_{\text{in}},
\end{equation}
where the new effective noise has correlators $\langle\tilde b_{\text{in}}(t)\tilde b_{\text{in}}(t')\rangle=n_{\text{thermal}}\delta(t-t')\times(\Gamma_{\mathrm m}/\Gamma_{\text{eff}})=n_{\text{thermal}}\delta(t-t')/\Lambda$,
where we have introduced a cooling parameter $\Lambda=\Gamma_{\text{eff}}/\Gamma_{\mathrm m}$.
In effect, we have modified parameters $\Gamma_{\text{eff}}=\Lambda \Gamma_{\text{m}}$ and $n_{\text{eff}}=n_{\text{thermal}}/\Lambda$, parametrize by $\Lambda$,
with the remainder of the Langevin equations unchanged.

The effect of cooling with an additional drive is illustrated for the DPSA in \cref{fig:cooling}.
In this figure, we show the plaquette with the extra cooling mode, and plot the gain and noise in both directions for \emph{fixed coupling constants}, but for varying levels of cooling $\Lambda$.
The reason for keeping the coupling rates rather than the cooperativities unchanged is that high cooperativities become unattainable for strongly broadened mechanical resonators.
We observe that the auxiliary cooling negatively affects gain, and forward added noise, which is due to the effective decrease in cooperativity.
On the other hand, the isolation bandwidth is increased, since it depends on the mechanical linewidth. 
Most importantly though, the backward noise can be strongly reduced.

\section{Stability}
\subsection{Theory}
It is important to know in which regimes the rotating-wave approximation (RWA) is valid.
In order to numerically analyze the system beyond the RWA, we use the Hamiltonian \cref{eq:H rotating frame}, but now keep all of the terms.
In the resulting Hamiltonian, there are two frequencies present, $\Omega_2+\delta_1$ and $\Omega_2+\delta_2$.
Our approach is taken from Ref.~\cite{SMalz2016}, but now extended to incorporate two frequencies. 
Note that other approaches are possible~\cite{SLi2015}.
In fact, we will choose the two frequencies to be $\tilde\Omega_1=\Omega_1+\delta_1+\Omega_2+\delta_2$ and $\tilde \Omega_2=\Omega_1+\delta_1-\Omega_2-\delta_2$, since it will lead to a more compact Floquet matrix.
If we restrict the theory to only the second Fourier frequency $\tilde \Omega_2$ we recover the theory from the two sections above.
While this does capture changes in effective parameters, it does not show instabilities that are present in the full description.
Collecting all system variables into a vector $\vec x=(b_1,b_1\dagg,b_2,b_2\dagg,d_1,d_1\dagg,d_2,d_2\dagg)^T$, we can write the equations of motion as
\begin{equation}
	\dot{\vec x}=\matr A(t)\vec x+\matr L\vec x_{\text{in}},
	\label{eq:full eom}
\end{equation}
where $A(t)$ is sometimes called Langevin matrix and
$\matr L=\diag(\sqrt{\Gamma_{\mathrm m,1}},\sqrt{\Gamma_{\mathrm m,1}},\sqrt{\Gamma_{\mathrm m,2}},\sqrt{\Gamma_{\mathrm m,2}},\sqrt{\kappa_1},\sqrt{\kappa_1},\sqrt{\kappa_2},\sqrt{\kappa_2})$.
Since the Hamiltonian has two Fourier frequencies, we can write the Langevin matrix in terms of its Fourier components as well
\begin{equation}
	\matr A(t)=\sum_{m,n}e^{im\tilde\Omega_1t+in\tilde\Omega_2t}\matr A^{(m,n)}.
\end{equation}
For the Hamiltonian we consider, the non-zero Fourier components $\matr A^{(m,n)}$ are those with $|m|,|n|\leq1$.
We can do the same to $\vec x$, which defines $\vec x^{(m,n)}(t)$. However the Fourier components of $\vec x$ will still contain fluctuations, so they are not time-independent.
We additionally perform a Fourier transform of all Fourier components of $\vec x$,
so that we can finally write the original equation of motion \eqref{eq:full eom} as ($\forall m,n$)
\begin{equation}
	i(\omega-\tilde\Omega_1m-\tilde\Omega_2n)\vec x^{(m,n)}(\omega)+\sum_{k=-1}^1\sum_{l=-1}^1\matr A^{(k,l)}\vec x^{(m-k,n-l)}(\omega)=-\delta_{m,0}\delta_{n,0}\matr L\vec x_{\text{in}}(\omega).
\end{equation}
While in principle there are infinitely many coupled equations (one for each combination of $m,n\in\Z$),
we have to truncate at a certain number to make the problem tractable. 
From the truncated matrix we calculate the scattering matrix between the Fourier modes.
Of particular interest here is the scattering between the zeroth Fourier modes, for we are interested in a signal on resonance.
Forward and reverse gain are defined in the same way as in the main text, except that they now refer to the equivalent elements
in relation to the scattering matrix between the zeroth Fourier modes. 

In general, we can distinguish four different regimes.
For sideband parameters $\Omega_i/\kappa$, $(\Omega_1-\Omega_2)/\kappa$ that are very large in comparison to the cooperativities $\cc_{ij}=G_{ij}/(\kappa_{i}\Omega_{j})$, the RWA theory is fully valid.
As the sideband parameters decrease, there is a regime where the effective parameters of the systems start to change.
Since isolation relies on fine-tuning parameters, it is very sensitive to such a change of parameters.
The RWA theory can be restored when working with renormalized parameters or when numerically optimizing the plaquette phase, as we demonstrate below.
As the nonresonant terms become even stronger, i.e., through increasing cooperativities or a lower sideband parameter, the system has a qualitatively different response.
Finally, there is a regime where the system invariably becomes unstable (cf.\ also Ref.~\cite{SLi2015}).

In the whole section we take $\kappa_1=\kappa_2=\kappa$ and $\Gamma_{\text{m},1}=\Gamma_{\text{m},2}=\Gamma_{\text{m}}=10^{-2}\kappa$ and choose units such that $\kappa=1$ for simplicity. 
Our proposed additional sideband cooling can increase $\Gamma_{\text{m}}$ beyond that value.
We note that a mechanical damping rate $\Gamma_{\text{m}}=\kappa$ quantitatively changes our conclusions below, since it enhances the detrimental effect of the counterrotating terms, but that the results are qualitatively the same.

\subsection{Optimizing coupling rates at finite sideband parameters}
Here we show by example (cf.~\cref{fig:nearRWA})  that for moderate sideband parameters and cooperativities, optimizing the plaquette phase leads to near ideal behavior.
\begin{figure}[ht]
	\centering
	\includegraphics[width=.32\textwidth]{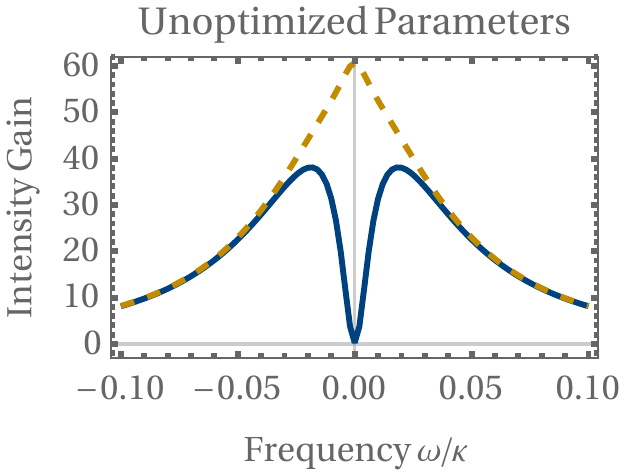}
	\includegraphics[width=.32\textwidth]{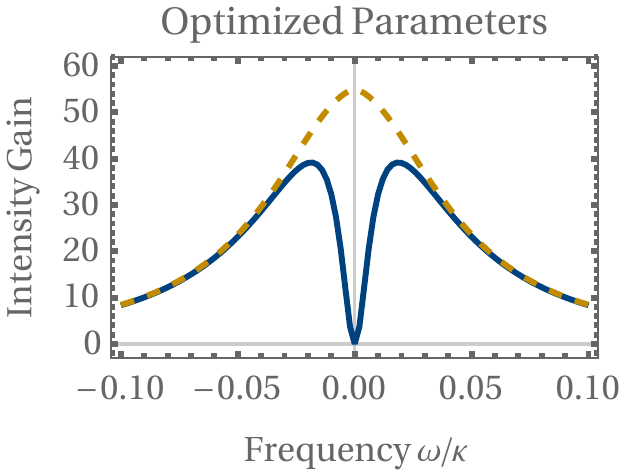}
	\includegraphics[width=.32\textwidth]{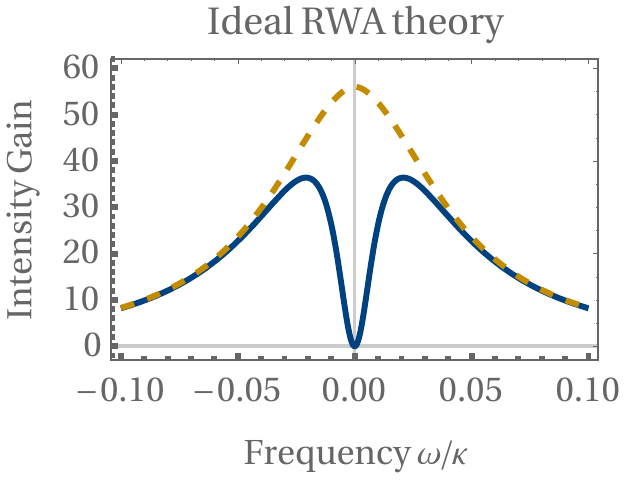}
	\caption{For finite sideband parameters, the RWA is not fully valid.
		The left panel shows forward and backward gain of the DPSA when using the coupling parameters of the ideal RWA theory.
		In the middle panel we have numerically optimized the plaquette phase.
		As comparison, the right panel shows the ideal theory.
		Parameters are $\cc_1=4,\cc_2=16,\Omega_1/\kappa=5,\Omega_2/\kappa=20$.
		Although the differences may appear subtle, we can quantify them (all on resonance): For unoptimized parameters, reverse gain is $0.23$, whereas forward gain is $61$. After optimizing, we have reverse gain of the order of $2\times10^{-17}$, which is essentially 0 within numerical errors, and forward gain 55.
		The ideal theory predicts 0 and 56.
	}
	\label{fig:nearRWA}
\end{figure}

\subsection{Qualitative and quantitative deviations from ideal theory}
For larger coupling rates we enter the regime where qualitative differences appear and the quantitative difference increase further. 
This is illustrated in \cref{fig:farRWA}.
Optimizing the plaquette phase still recovers isolation, as before, but forward gain takes a qualitatively different form and is considerably reduced in comparison with ideal theory.
\begin{figure}[ht]
	\centering
	\includegraphics[width=.32\textwidth]{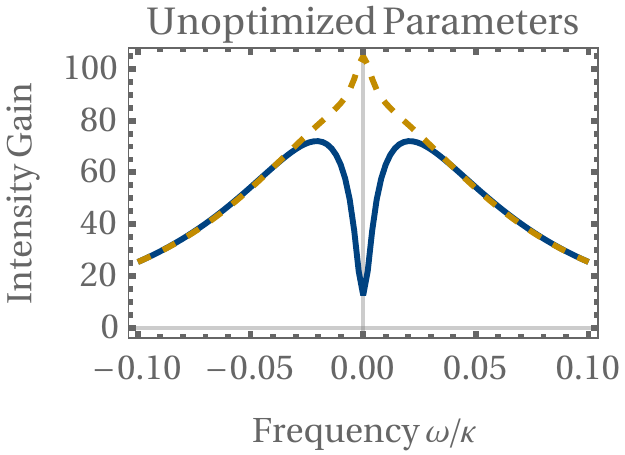}
	\includegraphics[width=.32\textwidth]{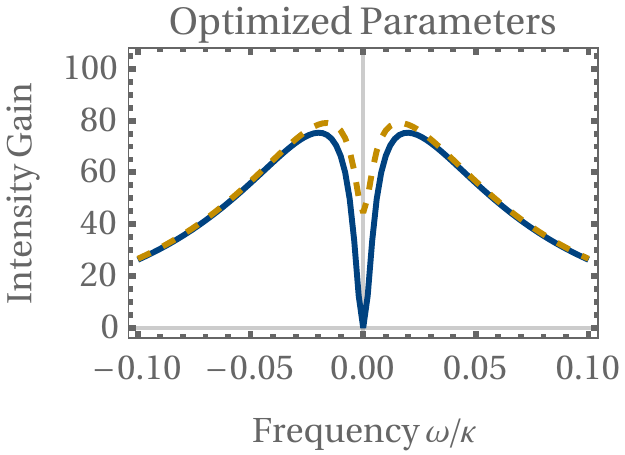}
	\includegraphics[width=.32\textwidth]{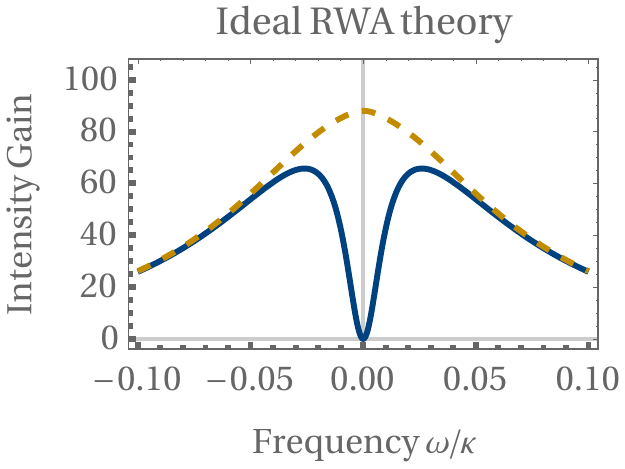}
	\caption{Increasing the coupling strength further, we enter the regime where qualitative differences appear.
		The left panel shows forward and backward gain of the DPSA when using the coupling parameters of the ideal RWA theory.
		In the middle panel we have numerically optimized the plaquette phase.
		As comparison, the right panel shows the ideal theory.
		Parameters are $\cc_1=6,\cc_2=36,\Omega_1/\kappa=5,\Omega_2/\kappa=20$.
		While in \cref{fig:nearRWA} the differences were only quantitative, here the forward gain in the middle panel shows a strong deviation from ideal theory (right panel).
		Again we list forward and reverse gain on resonance: unoptimized parameters, reverse/forward gain of $12.7$ and $105$, after optimizing, reverse/forward gain are $\sim8\times10^{-17}$ and $44.2$.
		As a comparison, the ideal theory predicts 0 and 88.
	}
	\label{fig:farRWA}
\end{figure}

\subsection{Instability threshold}
In this subsection we demonstrate a method to find the instability threshold.
We find that for a sideband parameter $\Omega/\kappa\gg1$ the eigenvectors are centered around a specific Fourier frequency and only mix with adjacent frequencies, due to the fact that the coupling is highly off-resonant. 
In the converse case, the eigenvectors are spread out over many Fourier frequencies. 
Thus, in the resolved sideband regime, we have to include only few Fourier frequencies.
Furthermore, due to the truncation at a certain Fourier index, there are eigenvectors that are concentrated at the edges of the Fourier domain that are not eigenvectors of the infinitely big matrix.
After solving the eigenvalue problem, we therefore select only those eigenvectors that have an appreciable support at the zeroth Fourier component.
We then look for the instability in this restricted set of modes.
We observe that there is an instability that occurs for certain cooperativities. For an analysis in a related system, see Ref.~\cite{SLi2015}.

We plot the instability threshold (in units of $\kappa$) as a function of resonator frequency in \cref{fig:resonator_frequency_dependence}. 
Note however that the plots in this section have been obtained for driving parameters given by the ideal theory.
We can make some progress through optimizing the driving parameters in the presence of the off-resonant terms,
as discussed in the previous subsections.
\begin{figure}[ht]
  \centering
  $\vcenter{\hbox{\includegraphics[width=.45\textwidth]{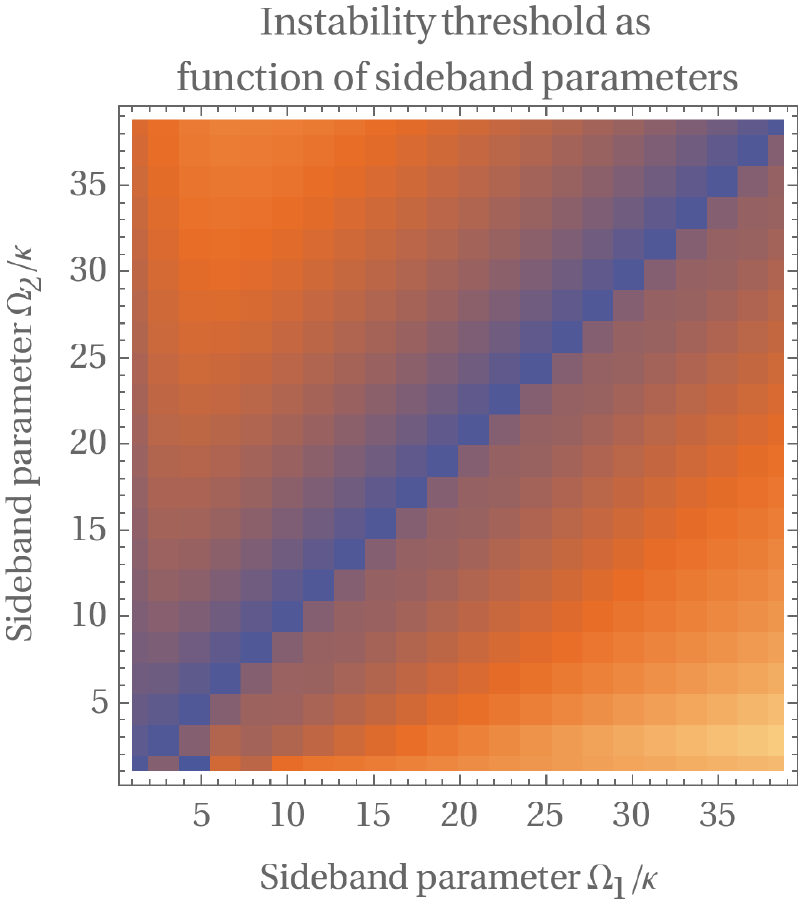}}}$
  \quad
  $\vcenter{\hbox{\includegraphics[width=.06\textwidth]{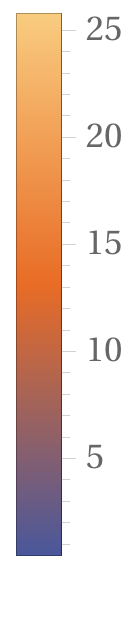}}}$
  \caption{We plot the cooperativity at which the instability occurs as a function of the two sideband parameters, $\Omega_1/\kappa$ and $\Omega_2/\kappa$.
  	The onset of stability is defined as the cooperativity $\cc_1$ (note $\cc_2=\cc_1^2$) at the real part of one of the eigenvalues becomes positive.
  	The set of eigenvalues is restricted to those belonging to eigenvectors with support at the zeroth Fourier component.
}
  \label{fig:resonator_frequency_dependence}
\end{figure}

\section{Connection to Metelmann and Clerk recipe for nonreciprocity \cite{SMetelmann2015}}
In this section we relate the calculations in the main text to the method for constructing nonreciprocal interactions presented in \cite{SMetelmann2015}.
Clearly, both conditions in \cref{eq:isolation condition} can only be fulfilled for complex susceptibilities, underscoring the importance of dissipation.
In order to distinguish coherent and dissipative parts of the coupling, we compare Eq.~(3) with the equations of motion for a nondegenerate parametric amplifier with interaction Hamiltonian
\begin{equation}
  H_{\text{int}}=\lambda a_1\dagg a_2\dagg+\text{H.c.},
  \label{eq:coherent coupling}
\end{equation}
namely
\begin{equation}
  \left( \frac{\kappa_1}{2}+\partial_t \right)\ev{\vec A_1}=i\mat{0&-\lambda\\\lambda^*&0}\ev{\vec A_2},\qquad
  \left( \frac{\kappa_2}{2}+\partial_t \right)\ev{\vec A_2}=i\mat{0&-\lambda\\\lambda^*&0}\ev{\vec A_1},
	\label{eq:NDPA eoms}
\end{equation}
where $\vec A_i\equiv (a_i,a_i\dagg)^T$.
The phase of $\lambda$ is arbitrary, as it is determined by the origin of time when going into the rotating frame.
\Cref{eq:NDPA eoms} suggests that coherent and dissipative interaction mediated by the mechanical resonators are the sum and difference of $iT_{12}$ and $iT_{21}$.
Thus, the coherent part is
\begin{equation}
	\lambda=\sum_i\Re[\chi_{\mathrm m,i}(0)]G_{i1}J_{i2}=\sum_i\frac{2\Gamma_{\mathrm m,i}G_{i1}J_{i2}}{\Gamma_{\mathrm m,i}^2+4\delta_i^2},
\end{equation}
whereas the dissipative part of the interaction is
\begin{equation}
	i\sigma=\sum_ii\Im[\chi_{\mathrm m,i}(0)]G_{i1}J_{i2}=i\sum_i\frac{4\delta_iG_{i1}J_{i2}}{\Gamma_{\mathrm m,i}^2+4\delta_i^2}.
\end{equation}
nonreciprocity is achieved if $\lambda+i\sigma=0$ but $\lambda,\sigma\neq0$, such that $\lambda-i\sigma\neq0$, a condition that the driving strengths Eq.~(5) in the main text fulfill.
For the DPPA, with Eq.~(5), and $\cc_1=1$ for simplicity, the equations of motion are 
\begin{subequations}
  \begin{align}
  	(\kappa_1+\partial_t)\ev{\vec A_1}&=i\mat{0&-\lambda-i\sigma\\\lambda^*-i\sigma^*&0}\ev{\vec A_2}=0,\\
  	\left[\frac{\kappa_2}{2}(1-\cc_2)+\partial_t\right]\ev{\vec A_2}&=i\mat{0&-\lambda+i\sigma\\\lambda^*+i\sigma^*&0}\ev{\vec A_1}
  	=2i\mat{0&\lambda\\-\lambda^*&0}\ev{\vec A_1},
  \end{align}
  \label{eq:pida eoms}
\end{subequations}
with 
\begin{equation}
  	\lambda=i\matr T_{12}(0)/2.
  \label{eq:pida lambda}
\end{equation}

Now that we have discerned which part of the interaction is dissipative and which is coherent,
we can map onto a quantum master equation.
Following \cite{SMetelmann2015}, the way to make a coherent interaction $H=Ja_1\dagg a_2\dagg+\text{H.c.}$ directional is 
by introducing a dissipative term in the QME of the form $\Gamma\mathcal L[z]\dens$, with
\begin{equation}
  z=\sqrt{2}(\cos\theta a_1+e^{i\varphi}\sin\theta a_2\dagg).
  \label{eq:dissipator}
\end{equation}
Directionality is obtained for the appropriate choice of $\theta$ and $\varphi$.
Indeed, we have 
\begin{subequations}
  \begin{align}
  	\ev{\dot a_1}&=-\left(\kappa_1/2+\Gamma\cos^2\theta\right)\ev{a_1}
  	+\left(iJ-\Gamma\sin\theta\cos\theta e^{i\varphi}\right)\ev{a_2\dagg},\\
  	\ev{\dot a_2\dagg}&=-\left(\kappa_2/2-\Gamma\sin^2\theta\right)\ev{a_2\dagg}
  	-\left(iJ^*-\Gamma\sin\theta\cos\theta e^{-i\varphi}\right)\ev{a_1}
  	.
  \end{align}
\end{subequations}
Setting $\Gamma$ and $\theta$ such that $\Gamma\sin\theta\cos\theta e^{i\varphi}=iJ$,
\begin{subequations}
  \begin{align}
  	\left(\frac{\kappa_1}{2}+|J\cot\theta|+\partial_t\right)\langle a_1\rangle&=0,\\
  	\left(\frac{\kappa_2}{2}-|J\tan\theta|+\partial_t\right)\langle a_2\dagg\rangle&=-2iJ^*\ev{a_1}.
  \end{align}
  \label{eq:QME eoms}
\end{subequations}
With $J=\lambda$, the RHS of \cref{eq:QME eoms} matches \cref{eq:pida eoms}.

In order to appropriately map the Langevin equations onto a master equation, we have to take into account that there are \emph{two} coherent interactions and \emph{two} baths.
Therefore, we have to repeat this procedure twice, making the two parts of the coherent interaction
\begin{subequations}
	\begin{align}
		i\lambda_1&=\Re[\chi_{\mathrm m,1}(0)]G_{11}J_{12}=\frac{\sqrt{\kappa_1\kappa_2\cc_2}}{4}e^{i\Phi/2},\\
		i\lambda_2&=\Re[\chi_{\mathrm m,2}(0)]G_{21}J_{22}=\frac{\sqrt{\kappa_1\kappa_2\cc_2}}{4}e^{-i\Phi/2}
	\end{align}
\end{subequations}
individually directional. This yields
\begin{subequations}
	\begin{align}
		\left( \frac{\kappa_1}{2}+\sum_i|\lambda_i\cot\theta_i|+\partial_t \right)\langle a_1\rangle&=0,\\
		\left( \frac{\kappa_2}{2}-\sum_i|\lambda_i\tan\theta_i|+\partial_t \right)\langle a_2\dagg\rangle&=-2(i\lambda_1^*+i\lambda_2^*)\langle a_1\rangle.
	\end{align}
\end{subequations}
Impedance matching gives $\tan\theta_i=4|\lambda_1|/\kappa_1$, and we recover \cref{eq:pida eoms}.

To illustrate why two baths are necessary, consider again \cref{eq:QME eoms}, which only has one bath.
Choosing $\theta$ such that $|\cot\theta|=\kappa_1/(2|J|)$ (impedance matching),
we do not quite recover \cref{eq:pida eoms}, but instead find 
\begin{equation}
	\frac{\kappa_2}{2}\left(1-\cc_2/2  \right)\langle a_2\dagg\rangle =-2i\lambda^*\langle a_1\rangle.
\end{equation}
Whilst this leads to the same interaction (RHS), the self-energies on the LHS differ.
Therefore, we need to theoretically include both baths in order to obtain an accurate representation of our system.
The reason for the factor of 2 difference lies in the fact that the coherent interactions differ by $\pi/2$ in phase, such that $|\lambda|=(|\lambda_1|+|\lambda_2|)/\sqrt{2}$ [cf.~\cref{eq:pida lambda}].

\bibliography{library}{}

\begin{thebibliography}{54}\makeatletter
\providecommand \@ifxundefined [1]{ \@ifx{#1\undefined}
}\providecommand \@ifnum [1]{ \ifnum #1\expandafter \@firstoftwo
 \else \expandafter \@secondoftwo
 \fi
}\providecommand \@ifx [1]{ \ifx #1\expandafter \@firstoftwo
 \else \expandafter \@secondoftwo
 \fi
}\providecommand \natexlab [1]{#1}\providecommand \enquote  [1]{``#1''}\providecommand \bibnamefont  [1]{#1}\providecommand \bibfnamefont [1]{#1}\providecommand \citenamefont [1]{#1}\providecommand \href@noop [0]{\@secondoftwo}\providecommand \href [0]{\begingroup \@sanitize@url \@href}\providecommand \@href[1]{\@@startlink{#1}\@@href}\providecommand \@@href[1]{\endgroup#1\@@endlink}\providecommand \@sanitize@url [0]{\catcode `\\12\catcode `\$12\catcode
  `\&12\catcode `\#12\catcode `\^12\catcode `\_12\catcode `\%12\relax}\providecommand \@@startlink[1]{}\providecommand \@@endlink[0]{}\providecommand \url  [0]{\begingroup\@sanitize@url \@url }\providecommand \@url [1]{\endgroup\@href {#1}{\urlprefix }}\providecommand \urlprefix  [0]{URL }\providecommand \Eprint [0]{\href }\providecommand \doibase [0]{http://dx.doi.org/}\providecommand \selectlanguage [0]{\@gobble}\providecommand \bibinfo  [0]{\@secondoftwo}\providecommand \bibfield  [0]{\@secondoftwo}\providecommand \translation [1]{[#1]}\providecommand \BibitemOpen [0]{}\providecommand \bibitemStop [0]{}\providecommand \bibitemNoStop [0]{.\EOS\space}\providecommand \EOS [0]{\spacefactor3000\relax}\providecommand \BibitemShut  [1]{\csname bibitem#1\endcsname}\let\auto@bib@innerbib\@empty
\bibitem [{\citenamefont {Gallo}\ \emph {et~al.}(2001)\citenamefont {Gallo},
  \citenamefont {Assanto}, \citenamefont {Parameswaran},\ and\ \citenamefont
  {Fejer}}]{Gallo2001}  \BibitemOpen
  \bibfield  {author} {\bibinfo {author} {\bibfnamefont {K.}~\bibnamefont
  {Gallo}}, \bibinfo {author} {\bibfnamefont {G.}~\bibnamefont {Assanto}},
  \bibinfo {author} {\bibfnamefont {K.~R.}\ \bibnamefont {Parameswaran}}, \
  and\ \bibinfo {author} {\bibfnamefont {M.~M.}\ \bibnamefont {Fejer}},\ }\href
  {\doibase 10.1063/1.1386407} {\bibfield  {journal} {\bibinfo  {journal}
  {Applied Physics Letters}\ }\textbf {\bibinfo {volume} {79}},\ \bibinfo
  {pages} {314} (\bibinfo {year} {2001})}\BibitemShut {NoStop}\bibitem [{\citenamefont {Koch}\ \emph {et~al.}(2010)\citenamefont {Koch},
  \citenamefont {Houck}, \citenamefont {Hur},\ and\ \citenamefont
  {Girvin}}]{Koch2010}  \BibitemOpen
  \bibfield  {author} {\bibinfo {author} {\bibfnamefont {J.}~\bibnamefont
  {Koch}}, \bibinfo {author} {\bibfnamefont {A.~A.}\ \bibnamefont {Houck}},
  \bibinfo {author} {\bibfnamefont {K.~L.}\ \bibnamefont {Hur}}, \ and\
  \bibinfo {author} {\bibfnamefont {S.~M.}\ \bibnamefont {Girvin}},\ }\href
  {\doibase 10.1103/PhysRevA.82.043811} {\bibfield  {journal} {\bibinfo
  {journal} {Physical Review A}\ }\textbf {\bibinfo {volume} {82}},\ \bibinfo
  {pages} {043811} (\bibinfo {year} {2010})}\BibitemShut {NoStop}\bibitem [{\citenamefont {Fang}\ \emph {et~al.}(2012)\citenamefont {Fang},
  \citenamefont {Yu},\ and\ \citenamefont {Fan}}]{Fang2012a}  \BibitemOpen
  \bibfield  {author} {\bibinfo {author} {\bibfnamefont {K.}~\bibnamefont
  {Fang}}, \bibinfo {author} {\bibfnamefont {Z.}~\bibnamefont {Yu}}, \ and\
  \bibinfo {author} {\bibfnamefont {S.}~\bibnamefont {Fan}},\ }\href {\doibase
  10.1038/nphoton.2012.236} {\bibfield  {journal} {\bibinfo  {journal} {Nature
  Photonics}\ }\textbf {\bibinfo {volume} {6}},\ \bibinfo {pages} {782}
  (\bibinfo {year} {2012})}\BibitemShut {NoStop}\bibitem [{\citenamefont {Sounas}\ \emph {et~al.}(2013)\citenamefont {Sounas},
  \citenamefont {Caloz},\ and\ \citenamefont {Al{\`{u}}}}]{Sounas2013}  \BibitemOpen
  \bibfield  {author} {\bibinfo {author} {\bibfnamefont {D.~L.}\ \bibnamefont
  {Sounas}}, \bibinfo {author} {\bibfnamefont {C.}~\bibnamefont {Caloz}}, \
  and\ \bibinfo {author} {\bibfnamefont {A.}~\bibnamefont {Al{\`{u}}}},\ }\href
  {\doibase 10.1038/ncomms3407} {\bibfield  {journal} {\bibinfo  {journal}
  {Nature Communications}\ }\textbf {\bibinfo {volume} {4}},\ \bibinfo {pages}
  {2407} (\bibinfo {year} {2013})},\ \Eprint {http://arxiv.org/abs/1011.1669}
  {arXiv:1011.1669} \BibitemShut {NoStop}\bibitem [{\citenamefont {Kamal}\ \emph {et~al.}(2011)\citenamefont {Kamal},
  \citenamefont {Clarke},\ and\ \citenamefont {Devoret}}]{Kamal2011}  \BibitemOpen
  \bibfield  {author} {\bibinfo {author} {\bibfnamefont {A.}~\bibnamefont
  {Kamal}}, \bibinfo {author} {\bibfnamefont {J.}~\bibnamefont {Clarke}}, \
  and\ \bibinfo {author} {\bibfnamefont {M.~H.}\ \bibnamefont {Devoret}},\
  }\href {\doibase 10.1038/nphys1893} {\bibfield  {journal} {\bibinfo
  {journal} {Nature Physics}\ }\textbf {\bibinfo {volume} {7}},\ \bibinfo
  {pages} {311} (\bibinfo {year} {2011})}\BibitemShut {NoStop}\bibitem [{\citenamefont {Hafezi}\ and\ \citenamefont
  {Rabl}(2012)}]{Hafezi2012}  \BibitemOpen
  \bibfield  {author} {\bibinfo {author} {\bibfnamefont {M.}~\bibnamefont
  {Hafezi}}\ and\ \bibinfo {author} {\bibfnamefont {P.}~\bibnamefont {Rabl}},\
  }\href {\doibase 10.1364/OE.20.007672} {\bibfield  {journal} {\bibinfo
  {journal} {Optics Express}\ }\textbf {\bibinfo {volume} {20}},\ \bibinfo
  {pages} {7672} (\bibinfo {year} {2012})},\ \Eprint
  {http://arxiv.org/abs/1107.3761} {arXiv:1107.3761} \BibitemShut {NoStop}\bibitem [{\citenamefont {Poulton}\ \emph {et~al.}(2012)\citenamefont
  {Poulton}, \citenamefont {Pant}, \citenamefont {Byrnes}, \citenamefont {Fan},
  \citenamefont {Steel},\ and\ \citenamefont {Eggleton}}]{Poulton2012}  \BibitemOpen
  \bibfield  {author} {\bibinfo {author} {\bibfnamefont {C.~G.}\ \bibnamefont
  {Poulton}}, \bibinfo {author} {\bibfnamefont {R.}~\bibnamefont {Pant}},
  \bibinfo {author} {\bibfnamefont {A.}~\bibnamefont {Byrnes}}, \bibinfo
  {author} {\bibfnamefont {S.}~\bibnamefont {Fan}}, \bibinfo {author}
  {\bibfnamefont {M.~J.}\ \bibnamefont {Steel}}, \ and\ \bibinfo {author}
  {\bibfnamefont {B.~J.}\ \bibnamefont {Eggleton}},\ }\href {\doibase
  10.1364/OE.20.021235} {\bibfield  {journal} {\bibinfo  {journal} {Optics
  Express}\ }\textbf {\bibinfo {volume} {20}},\ \bibinfo {pages} {21235}
  (\bibinfo {year} {2012})}\BibitemShut {NoStop}\bibitem [{\citenamefont {Longhi}(2013)}]{Longhi2013}  \BibitemOpen
  \bibfield  {author} {\bibinfo {author} {\bibfnamefont {S.}~\bibnamefont
  {Longhi}},\ }\href {\doibase 10.1364/OL.38.003570} {\bibfield  {journal}
  {\bibinfo  {journal} {Optics Letters}\ }\textbf {\bibinfo {volume} {38}},\
  \bibinfo {pages} {3570} (\bibinfo {year} {2013})},\ \Eprint
  {http://arxiv.org/abs/1310.5001v1} {arXiv:1310.5001v1} \BibitemShut {NoStop}\bibitem [{\citenamefont {Fang}\ \emph {et~al.}(2013)\citenamefont {Fang},
  \citenamefont {Yu},\ and\ \citenamefont {Fan}}]{Fang2013}  \BibitemOpen
  \bibfield  {author} {\bibinfo {author} {\bibfnamefont {K.}~\bibnamefont
  {Fang}}, \bibinfo {author} {\bibfnamefont {Z.}~\bibnamefont {Yu}}, \ and\
  \bibinfo {author} {\bibfnamefont {S.}~\bibnamefont {Fan}},\ }\href {\doibase
  10.1103/PhysRevB.87.060301} {\bibfield  {journal} {\bibinfo  {journal}
  {Physical Review B}\ }\textbf {\bibinfo {volume} {87}},\ \bibinfo {pages}
  {060301} (\bibinfo {year} {2013})}\BibitemShut {NoStop}\bibitem [{\citenamefont {Wang}\ \emph {et~al.}(2013)\citenamefont {Wang},
  \citenamefont {Zhou}, \citenamefont {Guo}, \citenamefont {Zhang},
  \citenamefont {Evers},\ and\ \citenamefont {Zhu}}]{Wang2013}  \BibitemOpen
  \bibfield  {author} {\bibinfo {author} {\bibfnamefont {D.-W.}\ \bibnamefont
  {Wang}}, \bibinfo {author} {\bibfnamefont {H.-T.}\ \bibnamefont {Zhou}},
  \bibinfo {author} {\bibfnamefont {M.-J.}\ \bibnamefont {Guo}}, \bibinfo
  {author} {\bibfnamefont {J.-X.}\ \bibnamefont {Zhang}}, \bibinfo {author}
  {\bibfnamefont {J.}~\bibnamefont {Evers}}, \ and\ \bibinfo {author}
  {\bibfnamefont {S.-Y.}\ \bibnamefont {Zhu}},\ }\href {\doibase
  10.1103/PhysRevLett.110.093901} {\bibfield  {journal} {\bibinfo  {journal}
  {Physical Review Letters}\ }\textbf {\bibinfo {volume} {110}},\ \bibinfo
  {pages} {093901} (\bibinfo {year} {2013})},\ \Eprint
  {http://arxiv.org/abs/1204.0837} {arXiv:1204.0837} \BibitemShut {NoStop}\bibitem [{\citenamefont {Abdo}\ \emph {et~al.}(2014)\citenamefont {Abdo},
  \citenamefont {Sliwa}, \citenamefont {Shankar}, \citenamefont {Hatridge},
  \citenamefont {Frunzio}, \citenamefont {Schoelkopf},\ and\ \citenamefont
  {Devoret}}]{Abdo2014}  \BibitemOpen
  \bibfield  {author} {\bibinfo {author} {\bibfnamefont {B.}~\bibnamefont
  {Abdo}}, \bibinfo {author} {\bibfnamefont {K.}~\bibnamefont {Sliwa}},
  \bibinfo {author} {\bibfnamefont {S.}~\bibnamefont {Shankar}}, \bibinfo
  {author} {\bibfnamefont {M.}~\bibnamefont {Hatridge}}, \bibinfo {author}
  {\bibfnamefont {L.}~\bibnamefont {Frunzio}}, \bibinfo {author} {\bibfnamefont
  {R.}~\bibnamefont {Schoelkopf}}, \ and\ \bibinfo {author} {\bibfnamefont
  {M.}~\bibnamefont {Devoret}},\ }\href {\doibase
  10.1103/PhysRevLett.112.167701} {\bibfield  {journal} {\bibinfo  {journal}
  {Physical Review Letters}\ }\textbf {\bibinfo {volume} {112}},\ \bibinfo
  {pages} {167701} (\bibinfo {year} {2014})},\ \Eprint
  {http://arxiv.org/abs/1311.5345} {arXiv:1311.5345} \BibitemShut {NoStop}\bibitem [{\citenamefont {Poshakinskiy}\ and\ \citenamefont
  {Poddubny}(2017)}]{Poshakinskiy2017}  \BibitemOpen
  \bibfield  {author} {\bibinfo {author} {\bibfnamefont {A.~V.}\ \bibnamefont
  {Poshakinskiy}}\ and\ \bibinfo {author} {\bibfnamefont {A.~N.}\ \bibnamefont
  {Poddubny}},\ }\href {\doibase 10.1103/PhysRevLett.118.156801} {\bibfield
  {journal} {\bibinfo  {journal} {Physical Review Letters}\ }\textbf {\bibinfo
  {volume} {118}},\ \bibinfo {pages} {156801} (\bibinfo {year}
  {2017})}\BibitemShut {NoStop}\bibitem [{\citenamefont {Sliwa}\ \emph {et~al.}(2015)\citenamefont {Sliwa},
  \citenamefont {Hatridge}, \citenamefont {Narla}, \citenamefont {Shankar},
  \citenamefont {Frunzio}, \citenamefont {Schoelkopf},\ and\ \citenamefont
  {Devoret}}]{Sliwa2015}  \BibitemOpen
  \bibfield  {author} {\bibinfo {author} {\bibfnamefont {K.~M.}\ \bibnamefont
  {Sliwa}}, \bibinfo {author} {\bibfnamefont {M.}~\bibnamefont {Hatridge}},
  \bibinfo {author} {\bibfnamefont {A.}~\bibnamefont {Narla}}, \bibinfo
  {author} {\bibfnamefont {S.}~\bibnamefont {Shankar}}, \bibinfo {author}
  {\bibfnamefont {L.}~\bibnamefont {Frunzio}}, \bibinfo {author} {\bibfnamefont
  {R.~J.}\ \bibnamefont {Schoelkopf}}, \ and\ \bibinfo {author} {\bibfnamefont
  {M.~H.}\ \bibnamefont {Devoret}},\ }\href {\doibase
  10.1103/PhysRevX.5.041020} {\bibfield  {journal} {\bibinfo  {journal}
  {Physical Review X}\ }\textbf {\bibinfo {volume} {5}},\ \bibinfo {pages}
  {041020} (\bibinfo {year} {2015})},\ \Eprint
  {http://arxiv.org/abs/1503.00209} {arXiv:1503.00209} \BibitemShut {NoStop}\bibitem [{\citenamefont {Lecocq}\ \emph {et~al.}(2017)\citenamefont {Lecocq},
  \citenamefont {Ranzani}, \citenamefont {Peterson}, \citenamefont {Cicak},
  \citenamefont {Simmonds}, \citenamefont {Teufel},\ and\ \citenamefont
  {Aumentado}}]{Lecocq2017}  \BibitemOpen
  \bibfield  {author} {\bibinfo {author} {\bibfnamefont {F.}~\bibnamefont
  {Lecocq}}, \bibinfo {author} {\bibfnamefont {L.}~\bibnamefont {Ranzani}},
  \bibinfo {author} {\bibfnamefont {G.~A.}\ \bibnamefont {Peterson}}, \bibinfo
  {author} {\bibfnamefont {K.}~\bibnamefont {Cicak}}, \bibinfo {author}
  {\bibfnamefont {R.~W.}\ \bibnamefont {Simmonds}}, \bibinfo {author}
  {\bibfnamefont {J.~D.}\ \bibnamefont {Teufel}}, \ and\ \bibinfo {author}
  {\bibfnamefont {J.}~\bibnamefont {Aumentado}},\ }\href {\doibase
  10.1103/PhysRevApplied.7.024028} {\bibfield  {journal} {\bibinfo  {journal}
  {Physical Review Applied}\ }\textbf {\bibinfo {volume} {7}},\ \bibinfo
  {pages} {024028} (\bibinfo {year} {2017})},\ \Eprint
  {http://arxiv.org/abs/1612.01438} {arXiv:1612.01438} \BibitemShut {NoStop}\bibitem [{\citenamefont {Abdo}\ \emph {et~al.}(2013)\citenamefont {Abdo},
  \citenamefont {Sliwa}, \citenamefont {Frunzio},\ and\ \citenamefont
  {Devoret}}]{Abdo2013}  \BibitemOpen
  \bibfield  {author} {\bibinfo {author} {\bibfnamefont {B.}~\bibnamefont
  {Abdo}}, \bibinfo {author} {\bibfnamefont {K.}~\bibnamefont {Sliwa}},
  \bibinfo {author} {\bibfnamefont {L.}~\bibnamefont {Frunzio}}, \ and\
  \bibinfo {author} {\bibfnamefont {M.}~\bibnamefont {Devoret}},\ }\href
  {\doibase 10.1103/PhysRevX.3.031001} {\bibfield  {journal} {\bibinfo
  {journal} {Physical Review X}\ }\textbf {\bibinfo {volume} {3}},\ \bibinfo
  {pages} {031001} (\bibinfo {year} {2013})},\ \Eprint
  {http://arxiv.org/abs/1302.4663} {arXiv:1302.4663} \BibitemShut {NoStop}\bibitem [{\citenamefont {Ranzani}\ and\ \citenamefont
  {Aumentado}(2014)}]{Ranzani2014}  \BibitemOpen
  \bibfield  {author} {\bibinfo {author} {\bibfnamefont {L.}~\bibnamefont
  {Ranzani}}\ and\ \bibinfo {author} {\bibfnamefont {J.}~\bibnamefont
  {Aumentado}},\ }\href
  {http://stacks.iop.org/1367-2630/16/i=10/a=103027?key=crossref.ed9cdef522ddb09357356fc11531b169}
  {\bibfield  {journal} {\bibinfo  {journal} {New Journal of Physics}\ }\textbf
  {\bibinfo {volume} {16}},\ \bibinfo {pages} {103027} (\bibinfo {year}
  {2014})},\ \Eprint {http://arxiv.org/abs/1406.4922} {arXiv:1406.4922}
  \BibitemShut {NoStop}\bibitem [{\citenamefont {Ranzani}\ and\ \citenamefont
  {Aumentado}(2015)}]{Ranzani2015}  \BibitemOpen
  \bibfield  {author} {\bibinfo {author} {\bibfnamefont {L.}~\bibnamefont
  {Ranzani}}\ and\ \bibinfo {author} {\bibfnamefont {J.}~\bibnamefont
  {Aumentado}},\ }\href {\doibase 10.1088/1367-2630/17/2/023024} {\bibfield
  {journal} {\bibinfo  {journal} {New Journal of Physics}\ }\textbf {\bibinfo
  {volume} {17}},\ \bibinfo {pages} {23024} (\bibinfo {year} {2015})},\ \Eprint
  {http://arxiv.org/abs/1406.4922v2} {arXiv:1406.4922v2} \BibitemShut {NoStop}\bibitem [{\citenamefont {Metelmann}\ and\ \citenamefont
  {Clerk}(2015)}]{Metelmann2015}  \BibitemOpen
  \bibfield  {author} {\bibinfo {author} {\bibfnamefont {A.}~\bibnamefont
  {Metelmann}}\ and\ \bibinfo {author} {\bibfnamefont {A.~A.}\ \bibnamefont
  {Clerk}},\ }\href {\doibase 10.1103/PhysRevX.5.021025} {\bibfield  {journal}
  {\bibinfo  {journal} {Physical Review X}\ }\textbf {\bibinfo {volume} {5}},\
  \bibinfo {pages} {021025} (\bibinfo {year} {2015})},\ \Eprint
  {http://arxiv.org/abs/1502.07274} {arXiv:1502.07274} \BibitemShut {NoStop}\bibitem [{\citenamefont {Fang}\ \emph {et~al.}(2017)\citenamefont {Fang},
  \citenamefont {Luo}, \citenamefont {Metelmann}, \citenamefont {Matheny},
  \citenamefont {Marquardt}, \citenamefont {Clerk},\ and\ \citenamefont
  {Painter}}]{Fang2017}  \BibitemOpen
  \bibfield  {author} {\bibinfo {author} {\bibfnamefont {K.}~\bibnamefont
  {Fang}}, \bibinfo {author} {\bibfnamefont {J.}~\bibnamefont {Luo}}, \bibinfo
  {author} {\bibfnamefont {A.}~\bibnamefont {Metelmann}}, \bibinfo {author}
  {\bibfnamefont {M.~H.}\ \bibnamefont {Matheny}}, \bibinfo {author}
  {\bibfnamefont {F.}~\bibnamefont {Marquardt}}, \bibinfo {author}
  {\bibfnamefont {A.~A.}\ \bibnamefont {Clerk}}, \ and\ \bibinfo {author}
  {\bibfnamefont {O.}~\bibnamefont {Painter}},\ }\href {\doibase
  10.1038/nphys4009} {\bibfield  {journal} {\bibinfo  {journal} {Nature
  Physics}\ }\textbf {\bibinfo {volume} {13}},\ \bibinfo {pages} {465}
  (\bibinfo {year} {2017})},\ \Eprint {http://arxiv.org/abs/1608.03620}
  {arXiv:1608.03620} \BibitemShut {NoStop}\bibitem [{\citenamefont {Kamal}\ and\ \citenamefont
  {Metelmann}(2017)}]{Kamal2016}  \BibitemOpen
  \bibfield  {author} {\bibinfo {author} {\bibfnamefont {A.}~\bibnamefont
  {Kamal}}\ and\ \bibinfo {author} {\bibfnamefont {A.}~\bibnamefont
  {Metelmann}},\ }\href {\doibase 10.1103/PhysRevApplied.7.034031} {\bibfield
  {journal} {\bibinfo  {journal} {Physical Review Applied}\ }\textbf {\bibinfo
  {volume} {7}},\ \bibinfo {pages} {034031} (\bibinfo {year} {2017})},\ \Eprint
  {http://arxiv.org/abs/1607.06822} {arXiv:1607.06822} \BibitemShut {NoStop}\bibitem [{\citenamefont {Metelmann}\ and\ \citenamefont
  {Clerk}(2017)}]{Metelmann2017}  \BibitemOpen
  \bibfield  {author} {\bibinfo {author} {\bibfnamefont {A.}~\bibnamefont
  {Metelmann}}\ and\ \bibinfo {author} {\bibfnamefont {A.~A.}\ \bibnamefont
  {Clerk}},\ }\href {\doibase 10.1103/PhysRevA.95.013837} {\bibfield  {journal}
  {\bibinfo  {journal} {Physical Review A}\ }\textbf {\bibinfo {volume} {95}},\
  \bibinfo {pages} {013837} (\bibinfo {year} {2017})},\ \Eprint
  {http://arxiv.org/abs/1610.06621} {arXiv:1610.06621} \BibitemShut {NoStop}\bibitem [{\citenamefont {Shen}\ \emph {et~al.}(2016)\citenamefont {Shen},
  \citenamefont {Zhang}, \citenamefont {Chen}, \citenamefont {Zou},
  \citenamefont {Xiao}, \citenamefont {Zou}, \citenamefont {Sun}, \citenamefont
  {Guo},\ and\ \citenamefont {Dong}}]{Shen2016}  \BibitemOpen
  \bibfield  {author} {\bibinfo {author} {\bibfnamefont {Z.}~\bibnamefont
  {Shen}}, \bibinfo {author} {\bibfnamefont {Y.-L.}\ \bibnamefont {Zhang}},
  \bibinfo {author} {\bibfnamefont {Y.}~\bibnamefont {Chen}}, \bibinfo {author}
  {\bibfnamefont {C.-L.}\ \bibnamefont {Zou}}, \bibinfo {author} {\bibfnamefont
  {Y.-F.}\ \bibnamefont {Xiao}}, \bibinfo {author} {\bibfnamefont {X.-B.}\
  \bibnamefont {Zou}}, \bibinfo {author} {\bibfnamefont {F.-W.}\ \bibnamefont
  {Sun}}, \bibinfo {author} {\bibfnamefont {G.-C.}\ \bibnamefont {Guo}}, \ and\
  \bibinfo {author} {\bibfnamefont {C.-H.}\ \bibnamefont {Dong}},\ }\href
  {\doibase 10.1038/nphoton.2016.161} {\bibfield  {journal} {\bibinfo
  {journal} {Nature Photonics}\ }\textbf {\bibinfo {volume} {10}},\ \bibinfo
  {pages} {657} (\bibinfo {year} {2016})},\ \Eprint
  {http://arxiv.org/abs/1604.02297} {arXiv:1604.02297} \BibitemShut {NoStop}\bibitem [{\citenamefont {Ruesink}\ \emph {et~al.}(2016)\citenamefont
  {Ruesink}, \citenamefont {Miri}, \citenamefont {Al{\`{u}}},\ and\
  \citenamefont {Verhagen}}]{Ruesink2016}  \BibitemOpen
  \bibfield  {author} {\bibinfo {author} {\bibfnamefont {F.}~\bibnamefont
  {Ruesink}}, \bibinfo {author} {\bibfnamefont {M.-A.}\ \bibnamefont {Miri}},
  \bibinfo {author} {\bibfnamefont {A.}~\bibnamefont {Al{\`{u}}}}, \ and\
  \bibinfo {author} {\bibfnamefont {E.}~\bibnamefont {Verhagen}},\ }\href
  {\doibase 10.1038/ncomms13662} {\bibfield  {journal} {\bibinfo  {journal}
  {Nature Communications}\ }\textbf {\bibinfo {volume} {7}},\ \bibinfo {pages}
  {13662} (\bibinfo {year} {2016})}\BibitemShut {NoStop}\bibitem [{\citenamefont {Xu}\ \emph {et~al.}(2016)\citenamefont {Xu},
  \citenamefont {Li}, \citenamefont {Chen},\ and\ \citenamefont
  {Liu}}]{Xu2016}  \BibitemOpen
  \bibfield  {author} {\bibinfo {author} {\bibfnamefont {X.-W.}\ \bibnamefont
  {Xu}}, \bibinfo {author} {\bibfnamefont {Y.}~\bibnamefont {Li}}, \bibinfo
  {author} {\bibfnamefont {A.-X.}\ \bibnamefont {Chen}}, \ and\ \bibinfo
  {author} {\bibfnamefont {Y.-X.}\ \bibnamefont {Liu}},\ }\href {\doibase
  10.1103/PhysRevA.93.023827} {\bibfield  {journal} {\bibinfo  {journal}
  {Physical Review A}\ }\textbf {\bibinfo {volume} {93}},\ \bibinfo {pages}
  {023827} (\bibinfo {year} {2016})}\BibitemShut {NoStop}\bibitem [{\citenamefont {Bernier}\ \emph {et~al.}(2017)\citenamefont
  {Bernier}, \citenamefont {T{\'{o}}th}, \citenamefont {Koottandavida},
  \citenamefont {Ioannou}, \citenamefont {Malz}, \citenamefont {Nunnenkamp},
  \citenamefont {Feofanov},\ and\ \citenamefont {Kippenberg}}]{Bernier2017}  \BibitemOpen
  \bibfield  {author} {\bibinfo {author} {\bibfnamefont {N.~R.}\ \bibnamefont
  {Bernier}}, \bibinfo {author} {\bibfnamefont {L.~D.}\ \bibnamefont
  {T{\'{o}}th}}, \bibinfo {author} {\bibfnamefont {A.}~\bibnamefont
  {Koottandavida}}, \bibinfo {author} {\bibfnamefont {M.~A.}\ \bibnamefont
  {Ioannou}}, \bibinfo {author} {\bibfnamefont {D.}~\bibnamefont {Malz}},
  \bibinfo {author} {\bibfnamefont {A.}~\bibnamefont {Nunnenkamp}}, \bibinfo
  {author} {\bibfnamefont {A.~K.}\ \bibnamefont {Feofanov}}, \ and\ \bibinfo
  {author} {\bibfnamefont {T.~J.}\ \bibnamefont {Kippenberg}},\ }\href
  {\doibase 10.1038/s41467-017-00447-1} {\bibfield  {journal} {\bibinfo
  {journal} {Nature Communications}\ }\textbf {\bibinfo {volume} {8}},\
  \bibinfo {pages} {604} (\bibinfo {year} {2017})},\ \Eprint
  {http://arxiv.org/abs/1612.08223} {arXiv:1612.08223} \BibitemShut {NoStop}\bibitem [{\citenamefont {Peterson}\ \emph {et~al.}(2017)\citenamefont
  {Peterson}, \citenamefont {Lecocq}, \citenamefont {Cicak}, \citenamefont
  {Simmonds}, \citenamefont {Aumentado},\ and\ \citenamefont
  {Teufel}}]{Peterson2017}  \BibitemOpen
  \bibfield  {author} {\bibinfo {author} {\bibfnamefont {G.~A.}\ \bibnamefont
  {Peterson}}, \bibinfo {author} {\bibfnamefont {F.}~\bibnamefont {Lecocq}},
  \bibinfo {author} {\bibfnamefont {K.}~\bibnamefont {Cicak}}, \bibinfo
  {author} {\bibfnamefont {R.~W.}\ \bibnamefont {Simmonds}}, \bibinfo {author}
  {\bibfnamefont {J.}~\bibnamefont {Aumentado}}, \ and\ \bibinfo {author}
  {\bibfnamefont {J.~D.}\ \bibnamefont {Teufel}},\ }\href {\doibase
  10.1103/PhysRevX.7.031001} {\bibfield  {journal} {\bibinfo  {journal}
  {Physical Review X}\ }\textbf {\bibinfo {volume} {7}},\ \bibinfo {pages}
  {031001} (\bibinfo {year} {2017})}\BibitemShut {NoStop}\bibitem [{\citenamefont {Tian}\ and\ \citenamefont {Li}(2017)}]{Tian2017}  \BibitemOpen
  \bibfield  {author} {\bibinfo {author} {\bibfnamefont {L.}~\bibnamefont
  {Tian}}\ and\ \bibinfo {author} {\bibfnamefont {Z.}~\bibnamefont {Li}},\
  }\href {\doibase 10.1103/PhysRevA.96.013808} {\bibfield  {journal} {\bibinfo
  {journal} {Physical Review A}\ }\textbf {\bibinfo {volume} {96}},\ \bibinfo
  {pages} {013808} (\bibinfo {year} {2017})}\BibitemShut {NoStop}\bibitem [{\citenamefont {Massel}\ \emph {et~al.}(2011)\citenamefont {Massel},
  \citenamefont {Heikkil{\"{a}}}, \citenamefont {Pirkkalainen}, \citenamefont
  {Cho}, \citenamefont {Saloniemi}, \citenamefont {Hakonen},\ and\
  \citenamefont {Sillanp{\"{a}}{\"{a}}}}]{Massel2011}  \BibitemOpen
  \bibfield  {author} {\bibinfo {author} {\bibfnamefont {F.}~\bibnamefont
  {Massel}}, \bibinfo {author} {\bibfnamefont {T.~T.}\ \bibnamefont
  {Heikkil{\"{a}}}}, \bibinfo {author} {\bibfnamefont {J.-M.}\ \bibnamefont
  {Pirkkalainen}}, \bibinfo {author} {\bibfnamefont {S.~U.}\ \bibnamefont
  {Cho}}, \bibinfo {author} {\bibfnamefont {H.}~\bibnamefont {Saloniemi}},
  \bibinfo {author} {\bibfnamefont {P.~J.}\ \bibnamefont {Hakonen}}, \ and\
  \bibinfo {author} {\bibfnamefont {M.~A.}\ \bibnamefont
  {Sillanp{\"{a}}{\"{a}}}},\ }\href {\doibase 10.1038/nature10628} {\bibfield
  {journal} {\bibinfo  {journal} {Nature}\ }\textbf {\bibinfo {volume} {480}},\
  \bibinfo {pages} {351} (\bibinfo {year} {2011})},\ \Eprint
  {http://arxiv.org/abs/1107.4903} {arXiv:1107.4903} \BibitemShut {NoStop}\bibitem [{\citenamefont {Metelmann}\ and\ \citenamefont
  {Clerk}(2014)}]{Metelmann2014}  \BibitemOpen
  \bibfield  {author} {\bibinfo {author} {\bibfnamefont {A.}~\bibnamefont
  {Metelmann}}\ and\ \bibinfo {author} {\bibfnamefont {A.~A.}\ \bibnamefont
  {Clerk}},\ }\href {\doibase 10.1103/PhysRevLett.112.133904} {\bibfield
  {journal} {\bibinfo  {journal} {Physical Review Letters}\ }\textbf {\bibinfo
  {volume} {112}},\ \bibinfo {pages} {133904} (\bibinfo {year}
  {2014})}\BibitemShut {NoStop}\bibitem [{\citenamefont {Nunnenkamp}\ \emph {et~al.}(2014)\citenamefont
  {Nunnenkamp}, \citenamefont {Sudhir}, \citenamefont {Feofanov}, \citenamefont
  {Roulet},\ and\ \citenamefont {Kippenberg}}]{Nunnenkamp2014}  \BibitemOpen
  \bibfield  {author} {\bibinfo {author} {\bibfnamefont {A.}~\bibnamefont
  {Nunnenkamp}}, \bibinfo {author} {\bibfnamefont {V.}~\bibnamefont {Sudhir}},
  \bibinfo {author} {\bibfnamefont {A.~K.}\ \bibnamefont {Feofanov}}, \bibinfo
  {author} {\bibfnamefont {A.}~\bibnamefont {Roulet}}, \ and\ \bibinfo {author}
  {\bibfnamefont {T.~J.}\ \bibnamefont {Kippenberg}},\ }\href {\doibase
  10.1103/PhysRevLett.113.023604} {\bibfield  {journal} {\bibinfo  {journal}
  {Physical Review Letters}\ }\textbf {\bibinfo {volume} {113}},\ \bibinfo
  {pages} {023604} (\bibinfo {year} {2014})}\BibitemShut {NoStop}\bibitem [{\citenamefont {T{\'{o}}th}\ \emph {et~al.}(2017)\citenamefont
  {T{\'{o}}th}, \citenamefont {Bernier}, \citenamefont {Nunnenkamp},
  \citenamefont {Feofanov},\ and\ \citenamefont {Kippenberg}}]{Toth2017}  \BibitemOpen
  \bibfield  {author} {\bibinfo {author} {\bibfnamefont {L.~D.}\ \bibnamefont
  {T{\'{o}}th}}, \bibinfo {author} {\bibfnamefont {N.~R.}\ \bibnamefont
  {Bernier}}, \bibinfo {author} {\bibfnamefont {A.}~\bibnamefont {Nunnenkamp}},
  \bibinfo {author} {\bibfnamefont {A.~K.}\ \bibnamefont {Feofanov}}, \ and\
  \bibinfo {author} {\bibfnamefont {T.~J.}\ \bibnamefont {Kippenberg}},\ }\href
  {\doibase 10.1038/nphys4121} {\bibfield  {journal} {\bibinfo  {journal}
  {Nature Physics}\ }\textbf {\bibinfo {volume} {13}},\ \bibinfo {pages} {787}
  (\bibinfo {year} {2017})}\BibitemShut {NoStop}\bibitem [{\citenamefont {Ockeloen-Korppi}\ \emph {et~al.}(2016)\citenamefont
  {Ockeloen-Korppi}, \citenamefont {Damsk{\"{a}}gg}, \citenamefont
  {Pirkkalainen}, \citenamefont {Heikkil{\"{a}}}, \citenamefont {Massel},\ and\
  \citenamefont {Sillanp{\"{a}}{\"{a}}}}]{Ockeloen-Korppi2016}  \BibitemOpen
  \bibfield  {author} {\bibinfo {author} {\bibfnamefont {C.~F.}\ \bibnamefont
  {Ockeloen-Korppi}}, \bibinfo {author} {\bibfnamefont {E.}~\bibnamefont
  {Damsk{\"{a}}gg}}, \bibinfo {author} {\bibfnamefont {J.-M.}\ \bibnamefont
  {Pirkkalainen}}, \bibinfo {author} {\bibfnamefont {T.~T.}\ \bibnamefont
  {Heikkil{\"{a}}}}, \bibinfo {author} {\bibfnamefont {F.}~\bibnamefont
  {Massel}}, \ and\ \bibinfo {author} {\bibfnamefont {M.~A.}\ \bibnamefont
  {Sillanp{\"{a}}{\"{a}}}},\ }\href {\doibase 10.1103/PhysRevX.6.041024}
  {\bibfield  {journal} {\bibinfo  {journal} {Physical Review X}\ }\textbf
  {\bibinfo {volume} {6}},\ \bibinfo {pages} {041024} (\bibinfo {year}
  {2016})},\ \Eprint {http://arxiv.org/abs/1602.05779} {arXiv:1602.05779}
  \BibitemShut {NoStop}\bibitem [{\citenamefont {Suh}\ \emph {et~al.}(2014)\citenamefont {Suh},
  \citenamefont {Weinstein}, \citenamefont {Lei}, \citenamefont {Wollman},
  \citenamefont {Steinke}, \citenamefont {Meystre}, \citenamefont {Clerk},\
  and\ \citenamefont {Schwab}}]{Suh2014}  \BibitemOpen
  \bibfield  {author} {\bibinfo {author} {\bibfnamefont {J.}~\bibnamefont
  {Suh}}, \bibinfo {author} {\bibfnamefont {A.~J.}\ \bibnamefont {Weinstein}},
  \bibinfo {author} {\bibfnamefont {C.~U.}\ \bibnamefont {Lei}}, \bibinfo
  {author} {\bibfnamefont {E.~E.}\ \bibnamefont {Wollman}}, \bibinfo {author}
  {\bibfnamefont {S.~K.}\ \bibnamefont {Steinke}}, \bibinfo {author}
  {\bibfnamefont {P.}~\bibnamefont {Meystre}}, \bibinfo {author} {\bibfnamefont
  {A.~A.}\ \bibnamefont {Clerk}}, \ and\ \bibinfo {author} {\bibfnamefont
  {K.~C.}\ \bibnamefont {Schwab}},\ }\href {\doibase 10.1126/science.1253258}
  {\bibfield  {journal} {\bibinfo  {journal} {Science}\ }\textbf {\bibinfo
  {volume} {344}},\ \bibinfo {pages} {1262} (\bibinfo {year} {2014})},\ \Eprint
  {http://arxiv.org/abs/1312.4084} {arXiv:1312.4084} \BibitemShut {NoStop}\bibitem [{\citenamefont {Pirkkalainen}\ \emph {et~al.}(2015)\citenamefont
  {Pirkkalainen}, \citenamefont {Damsk{\"{a}}gg}, \citenamefont {Brandt},
  \citenamefont {Massel},\ and\ \citenamefont
  {Sillanp{\"{a}}{\"{a}}}}]{Pirkkalainen2015}  \BibitemOpen
  \bibfield  {author} {\bibinfo {author} {\bibfnamefont {J.-M.}\ \bibnamefont
  {Pirkkalainen}}, \bibinfo {author} {\bibfnamefont {E.}~\bibnamefont
  {Damsk{\"{a}}gg}}, \bibinfo {author} {\bibfnamefont {M.}~\bibnamefont
  {Brandt}}, \bibinfo {author} {\bibfnamefont {F.}~\bibnamefont {Massel}}, \
  and\ \bibinfo {author} {\bibfnamefont {M.~A.}\ \bibnamefont
  {Sillanp{\"{a}}{\"{a}}}},\ }\href {\doibase 10.1103/PhysRevLett.115.243601}
  {\bibfield  {journal} {\bibinfo  {journal} {Physical Review Letters}\
  }\textbf {\bibinfo {volume} {115}},\ \bibinfo {pages} {243601} (\bibinfo
  {year} {2015})}\BibitemShut {NoStop}\bibitem [{\citenamefont {Wollman}\ \emph {et~al.}(2015)\citenamefont
  {Wollman}, \citenamefont {Lei}, \citenamefont {Weinstein}, \citenamefont
  {Suh}, \citenamefont {Kronwald}, \citenamefont {Marquardt}, \citenamefont
  {Clerk},\ and\ \citenamefont {Schwab}}]{Wollman2015}  \BibitemOpen
  \bibfield  {author} {\bibinfo {author} {\bibfnamefont {E.~E.}\ \bibnamefont
  {Wollman}}, \bibinfo {author} {\bibfnamefont {C.~U.}\ \bibnamefont {Lei}},
  \bibinfo {author} {\bibfnamefont {A.~J.}\ \bibnamefont {Weinstein}}, \bibinfo
  {author} {\bibfnamefont {J.}~\bibnamefont {Suh}}, \bibinfo {author}
  {\bibfnamefont {A.}~\bibnamefont {Kronwald}}, \bibinfo {author}
  {\bibfnamefont {F.}~\bibnamefont {Marquardt}}, \bibinfo {author}
  {\bibfnamefont {A.~A.}\ \bibnamefont {Clerk}}, \ and\ \bibinfo {author}
  {\bibfnamefont {K.~C.}\ \bibnamefont {Schwab}},\ }\href {\doibase
  10.1126/science.aac5138} {\bibfield  {journal} {\bibinfo  {journal}
  {Science}\ }\textbf {\bibinfo {volume} {349}},\ \bibinfo {pages} {952}
  (\bibinfo {year} {2015})},\ \Eprint {http://arxiv.org/abs/1507.01662}
  {arXiv:1507.01662} \BibitemShut {NoStop}\bibitem [{\citenamefont {Lei}\ \emph {et~al.}(2016)\citenamefont {Lei},
  \citenamefont {Weinstein}, \citenamefont {Suh}, \citenamefont {Wollman},
  \citenamefont {Kronwald}, \citenamefont {Marquardt}, \citenamefont {Clerk},\
  and\ \citenamefont {Schwab}}]{Lei2016}  \BibitemOpen
  \bibfield  {author} {\bibinfo {author} {\bibfnamefont {C.~U.}\ \bibnamefont
  {Lei}}, \bibinfo {author} {\bibfnamefont {A.~J.}\ \bibnamefont {Weinstein}},
  \bibinfo {author} {\bibfnamefont {J.}~\bibnamefont {Suh}}, \bibinfo {author}
  {\bibfnamefont {E.~E.}\ \bibnamefont {Wollman}}, \bibinfo {author}
  {\bibfnamefont {A.}~\bibnamefont {Kronwald}}, \bibinfo {author}
  {\bibfnamefont {F.}~\bibnamefont {Marquardt}}, \bibinfo {author}
  {\bibfnamefont {A.~A.}\ \bibnamefont {Clerk}}, \ and\ \bibinfo {author}
  {\bibfnamefont {K.~C.}\ \bibnamefont {Schwab}},\ }\href {\doibase
  10.1103/PhysRevLett.117.100801} {\bibfield  {journal} {\bibinfo  {journal}
  {Physical Review Letters}\ }\textbf {\bibinfo {volume} {117}},\ \bibinfo
  {pages} {100801} (\bibinfo {year} {2016})},\ \Eprint
  {http://arxiv.org/abs/1605.08148} {arXiv:1605.08148} \BibitemShut {NoStop}\bibitem [{\citenamefont {Clerk}\ \emph {et~al.}(2010)\citenamefont {Clerk},
  \citenamefont {Devoret}, \citenamefont {Girvin}, \citenamefont {Marquardt},\
  and\ \citenamefont {Schoelkopf}}]{Clerk2010}  \BibitemOpen
  \bibfield  {author} {\bibinfo {author} {\bibfnamefont {A.~A.}\ \bibnamefont
  {Clerk}}, \bibinfo {author} {\bibfnamefont {M.~H.}\ \bibnamefont {Devoret}},
  \bibinfo {author} {\bibfnamefont {S.~M.}\ \bibnamefont {Girvin}}, \bibinfo
  {author} {\bibfnamefont {F.}~\bibnamefont {Marquardt}}, \ and\ \bibinfo
  {author} {\bibfnamefont {R.~J.}\ \bibnamefont {Schoelkopf}},\ }\href
  {\doibase 10.1103/RevModPhys.82.1155} {\bibfield  {journal} {\bibinfo
  {journal} {Reviews of Modern Physics}\ }\textbf {\bibinfo {volume} {82}},\
  \bibinfo {pages} {1155} (\bibinfo {year} {2010})}\BibitemShut {NoStop}\bibitem [{\citenamefont {Aspelmeyer}\ \emph {et~al.}(2014)\citenamefont
  {Aspelmeyer}, \citenamefont {Kippenberg},\ and\ \citenamefont
  {Marquardt}}]{Aspelmeyer2014}  \BibitemOpen
  \bibfield  {author} {\bibinfo {author} {\bibfnamefont {M.}~\bibnamefont
  {Aspelmeyer}}, \bibinfo {author} {\bibfnamefont {T.~J.}\ \bibnamefont
  {Kippenberg}}, \ and\ \bibinfo {author} {\bibfnamefont {F.}~\bibnamefont
  {Marquardt}},\ }\href {\doibase 10.1103/RevModPhys.86.1391} {\bibfield
  {journal} {\bibinfo  {journal} {Reviews of Modern Physics}\ }\textbf
  {\bibinfo {volume} {86}},\ \bibinfo {pages} {1391} (\bibinfo {year}
  {2014})}\BibitemShut {NoStop}\bibitem [{\citenamefont {Gardiner}\ and\ \citenamefont
  {Collett}(1985)}]{Gardiner1985}  \BibitemOpen
  \bibfield  {author} {\bibinfo {author} {\bibfnamefont {C.~W.}\ \bibnamefont
  {Gardiner}}\ and\ \bibinfo {author} {\bibfnamefont {M.~J.}\ \bibnamefont
  {Collett}},\ }\href {\doibase 10.1103/PhysRevA.31.3761} {\bibfield  {journal}
  {\bibinfo  {journal} {Physical Review A}\ }\textbf {\bibinfo {volume} {31}},\
  \bibinfo {pages} {3761} (\bibinfo {year} {1985})}\BibitemShut {NoStop}\bibitem [{\citenamefont {Gardiner}\ and\ \citenamefont
  {Zoller}(2004)}]{gardiner2004quantum}  \BibitemOpen
  \bibfield  {author} {\bibinfo {author} {\bibfnamefont {C.}~\bibnamefont
  {Gardiner}}\ and\ \bibinfo {author} {\bibfnamefont {P.}~\bibnamefont
  {Zoller}},\ }\href {https://books.google.de/books?id=a{\_}xsT8oGhdgC} {\emph
  {\bibinfo {title} {{Quantum Noise: A Handbook of Markovian and Non-Markovian
  Quantum Stochastic Methods with Applications to Quantum Optics}}}},\ Springer
  Series in Synergetics\ (\bibinfo  {publisher} {Springer},\ \bibinfo {year}
  {2004})\BibitemShut {NoStop}\bibitem [{Note1()}]{Note1}  \BibitemOpen
  \bibinfo {note} {This condition causes the moduli of the transmission
  amplitudes via resonator 1 and 2 to coincide, $|G_{11}J_{21}\chi _{\protect
  \mathrm m,1}(0)|=|G_{12}J_{22}\chi _{\protect \mathrm m,2}(0)|$, needed for
  complete destructive interference.}\BibitemShut {Stop}\bibitem [{Note2()}]{Note2}  \BibitemOpen
  \bibinfo {note} {Private communication with Anja Metelmann. Note that
  choosing $G_{21}=J_{22}=0$, $G_{11}=J_{12}$, $\delta =0$, yields reciprocal
  amplifier with unlimited gain-bandwidth product~\cite
  {Metelmann2014}.}\BibitemShut {Stop}\bibitem [{\citenamefont {Caves}(1982)}]{Caves1982}  \BibitemOpen
  \bibfield  {author} {\bibinfo {author} {\bibfnamefont {C.~M.}\ \bibnamefont
  {Caves}},\ }\href {\doibase 10.1103/PhysRevD.26.1817} {\bibfield  {journal}
  {\bibinfo  {journal} {Physical Review D}\ }\textbf {\bibinfo {volume} {26}},\
  \bibinfo {pages} {1817} (\bibinfo {year} {1982})}\BibitemShut {NoStop}\bibitem [{\citenamefont {Braginsky}\ \emph {et~al.}(1980)\citenamefont
  {Braginsky}, \citenamefont {Vorontsov},\ and\ \citenamefont
  {Thorne}}]{Braginsky1980}  \BibitemOpen
  \bibfield  {author} {\bibinfo {author} {\bibfnamefont {V.~B.}\ \bibnamefont
  {Braginsky}}, \bibinfo {author} {\bibfnamefont {Y.~I.}\ \bibnamefont
  {Vorontsov}}, \ and\ \bibinfo {author} {\bibfnamefont {K.~S.}\ \bibnamefont
  {Thorne}},\ }\href {\doibase 10.1126/science.209.4456.547} {\bibfield
  {journal} {\bibinfo  {journal} {Science}\ }\textbf {\bibinfo {volume}
  {209}},\ \bibinfo {pages} {547} (\bibinfo {year} {1980})}\BibitemShut
  {NoStop}\bibitem [{\citenamefont {Clerk}\ \emph {et~al.}(2008)\citenamefont {Clerk},
  \citenamefont {Marquardt},\ and\ \citenamefont {Jacobs}}]{Clerk2008}  \BibitemOpen
  \bibfield  {author} {\bibinfo {author} {\bibfnamefont {A.~A.}\ \bibnamefont
  {Clerk}}, \bibinfo {author} {\bibfnamefont {F.}~\bibnamefont {Marquardt}}, \
  and\ \bibinfo {author} {\bibfnamefont {K.}~\bibnamefont {Jacobs}},\ }\href
  {\doibase 10.1088/1367-2630/10/9/095010} {\bibfield  {journal} {\bibinfo
  {journal} {New Journal of Physics}\ }\textbf {\bibinfo {volume} {10}},\
  \bibinfo {pages} {095010} (\bibinfo {year} {2008})},\ \Eprint
  {http://arxiv.org/abs/0802.1842} {arXiv:0802.1842} \BibitemShut {NoStop}\bibitem [{\citenamefont {Hertzberg}\ \emph {et~al.}(2010)\citenamefont
  {Hertzberg}, \citenamefont {Rocheleau}, \citenamefont {Ndukum}, \citenamefont
  {Savva}, \citenamefont {Clerk},\ and\ \citenamefont
  {Schwab}}]{Hertzberg2010}  \BibitemOpen
  \bibfield  {author} {\bibinfo {author} {\bibfnamefont {J.~B.}\ \bibnamefont
  {Hertzberg}}, \bibinfo {author} {\bibfnamefont {T.}~\bibnamefont
  {Rocheleau}}, \bibinfo {author} {\bibfnamefont {T.}~\bibnamefont {Ndukum}},
  \bibinfo {author} {\bibfnamefont {M.}~\bibnamefont {Savva}}, \bibinfo
  {author} {\bibfnamefont {A.~A.}\ \bibnamefont {Clerk}}, \ and\ \bibinfo
  {author} {\bibfnamefont {K.~C.}\ \bibnamefont {Schwab}},\ }\href {\doibase
  10.1038/nphys1479} {\bibfield  {journal} {\bibinfo  {journal} {Nature
  Physics}\ }\textbf {\bibinfo {volume} {6}},\ \bibinfo {pages} {213} (\bibinfo
  {year} {2010})}\BibitemShut {NoStop}\bibitem [{\citenamefont {Woolley}\ and\ \citenamefont
  {Clerk}(2013)}]{Woolley2013}  \BibitemOpen
  \bibfield  {author} {\bibinfo {author} {\bibfnamefont {M.~J.}\ \bibnamefont
  {Woolley}}\ and\ \bibinfo {author} {\bibfnamefont {A.~A.}\ \bibnamefont
  {Clerk}},\ }\href {\doibase 10.1103/PhysRevA.87.063846} {\bibfield  {journal}
  {\bibinfo  {journal} {Physical Review A}\ }\textbf {\bibinfo {volume} {87}},\
  \bibinfo {pages} {063846} (\bibinfo {year} {2013})},\ \Eprint
  {http://arxiv.org/abs/1304.4059} {arXiv:1304.4059} \BibitemShut {NoStop}\bibitem [{\citenamefont {Malz}\ and\ \citenamefont
  {Nunnenkamp}(2016{\natexlab{a}})}]{Malz2016}  \BibitemOpen
  \bibfield  {author} {\bibinfo {author} {\bibfnamefont {D.}~\bibnamefont
  {Malz}}\ and\ \bibinfo {author} {\bibfnamefont {A.}~\bibnamefont
  {Nunnenkamp}},\ }\href {\doibase 10.1103/PhysRevA.94.023803} {\bibfield
  {journal} {\bibinfo  {journal} {Physical Review A}\ }\textbf {\bibinfo
  {volume} {94}},\ \bibinfo {pages} {023803} (\bibinfo {year}
  {2016}{\natexlab{a}})},\ \Eprint {http://arxiv.org/abs/1605.04749}
  {arXiv:1605.04749} \BibitemShut {NoStop}\bibitem [{\citenamefont {Malz}\ and\ \citenamefont
  {Nunnenkamp}(2016{\natexlab{b}})}]{Malz2016b}  \BibitemOpen
  \bibfield  {author} {\bibinfo {author} {\bibfnamefont {D.}~\bibnamefont
  {Malz}}\ and\ \bibinfo {author} {\bibfnamefont {A.}~\bibnamefont
  {Nunnenkamp}},\ }\href {\doibase 10.1103/PhysRevA.94.053820} {\bibfield
  {journal} {\bibinfo  {journal} {Physical Review A}\ }\textbf {\bibinfo
  {volume} {94}},\ \bibinfo {pages} {053820} (\bibinfo {year}
  {2016}{\natexlab{b}})},\ \Eprint {http://arxiv.org/abs/1610.00154}
  {arXiv:1610.00154} \BibitemShut {NoStop}\bibitem [{\citenamefont {Li}\ \emph {et~al.}(2015)\citenamefont {Li},
  \citenamefont {Haghighi}, \citenamefont {Malossi}, \citenamefont {Zippilli},\
  and\ \citenamefont {Vitali}}]{Li2015}  \BibitemOpen
  \bibfield  {author} {\bibinfo {author} {\bibfnamefont {J.}~\bibnamefont
  {Li}}, \bibinfo {author} {\bibfnamefont {I.~M.}\ \bibnamefont {Haghighi}},
  \bibinfo {author} {\bibfnamefont {N.}~\bibnamefont {Malossi}}, \bibinfo
  {author} {\bibfnamefont {S.}~\bibnamefont {Zippilli}}, \ and\ \bibinfo
  {author} {\bibfnamefont {D.}~\bibnamefont {Vitali}},\ }\href {\doibase
  10.1088/1367-2630/17/10/103037} {\bibfield  {journal} {\bibinfo  {journal}
  {New Journal of Physics}\ }\textbf {\bibinfo {volume} {17}},\ \bibinfo
  {pages} {103037} (\bibinfo {year} {2015})}\BibitemShut {NoStop}\bibitem [{\citenamefont {Teufel}\ \emph
  {et~al.}(2011{\natexlab{a}})\citenamefont {Teufel}, \citenamefont {Li},
  \citenamefont {Allman}, \citenamefont {Cicak}, \citenamefont {Sirois},
  \citenamefont {Whittaker},\ and\ \citenamefont {Simmonds}}]{Teufel2011a}  \BibitemOpen
  \bibfield  {author} {\bibinfo {author} {\bibfnamefont {J.~D.}\ \bibnamefont
  {Teufel}}, \bibinfo {author} {\bibfnamefont {D.}~\bibnamefont {Li}}, \bibinfo
  {author} {\bibfnamefont {M.~S.}\ \bibnamefont {Allman}}, \bibinfo {author}
  {\bibfnamefont {K.}~\bibnamefont {Cicak}}, \bibinfo {author} {\bibfnamefont
  {A.~J.}\ \bibnamefont {Sirois}}, \bibinfo {author} {\bibfnamefont {J.~D.}\
  \bibnamefont {Whittaker}}, \ and\ \bibinfo {author} {\bibfnamefont {R.~W.}\
  \bibnamefont {Simmonds}},\ }\href {\doibase 10.1038/nature09898} {\bibfield
  {journal} {\bibinfo  {journal} {Nature}\ }\textbf {\bibinfo {volume} {471}},\
  \bibinfo {pages} {204} (\bibinfo {year} {2011}{\natexlab{a}})}\BibitemShut
  {NoStop}\bibitem [{\citenamefont {Teufel}\ \emph
  {et~al.}(2011{\natexlab{b}})\citenamefont {Teufel}, \citenamefont {Donner},
  \citenamefont {Li}, \citenamefont {Harlow}, \citenamefont {Allman},
  \citenamefont {Cicak}, \citenamefont {Sirois}, \citenamefont {Whittaker},
  \citenamefont {Lehnert},\ and\ \citenamefont {Simmonds}}]{Teufel2011}  \BibitemOpen
  \bibfield  {author} {\bibinfo {author} {\bibfnamefont {J.~D.}\ \bibnamefont
  {Teufel}}, \bibinfo {author} {\bibfnamefont {T.}~\bibnamefont {Donner}},
  \bibinfo {author} {\bibfnamefont {D.}~\bibnamefont {Li}}, \bibinfo {author}
  {\bibfnamefont {J.~W.}\ \bibnamefont {Harlow}}, \bibinfo {author}
  {\bibfnamefont {M.~S.}\ \bibnamefont {Allman}}, \bibinfo {author}
  {\bibfnamefont {K.}~\bibnamefont {Cicak}}, \bibinfo {author} {\bibfnamefont
  {A.~J.}\ \bibnamefont {Sirois}}, \bibinfo {author} {\bibfnamefont {J.~D.}\
  \bibnamefont {Whittaker}}, \bibinfo {author} {\bibfnamefont {K.~W.}\
  \bibnamefont {Lehnert}}, \ and\ \bibinfo {author} {\bibfnamefont {R.~W.}\
  \bibnamefont {Simmonds}},\ }\href {\doibase 10.1038/nature10261} {\bibfield
  {journal} {\bibinfo  {journal} {Nature}\ }\textbf {\bibinfo {volume} {475}},\
  \bibinfo {pages} {359} (\bibinfo {year} {2011}{\natexlab{b}})}\BibitemShut
  {NoStop}\bibitem [{\citenamefont {Chan}\ \emph {et~al.}(2011)\citenamefont {Chan},
  \citenamefont {Alegre}, \citenamefont {Safavi-Naeini}, \citenamefont {Hill},
  \citenamefont {Krause}, \citenamefont {Gr{\"{o}}blacher}, \citenamefont
  {Aspelmeyer},\ and\ \citenamefont {Painter}}]{Chan2011}  \BibitemOpen
  \bibfield  {author} {\bibinfo {author} {\bibfnamefont {J.}~\bibnamefont
  {Chan}}, \bibinfo {author} {\bibfnamefont {T.~P.~M.}\ \bibnamefont {Alegre}},
  \bibinfo {author} {\bibfnamefont {A.~H.}\ \bibnamefont {Safavi-Naeini}},
  \bibinfo {author} {\bibfnamefont {J.~T.}\ \bibnamefont {Hill}}, \bibinfo
  {author} {\bibfnamefont {A.}~\bibnamefont {Krause}}, \bibinfo {author}
  {\bibfnamefont {S.}~\bibnamefont {Gr{\"{o}}blacher}}, \bibinfo {author}
  {\bibfnamefont {M.}~\bibnamefont {Aspelmeyer}}, \ and\ \bibinfo {author}
  {\bibfnamefont {O.}~\bibnamefont {Painter}},\ }\href {\doibase
  10.1038/nature10461} {\bibfield  {journal} {\bibinfo  {journal} {Nature}\
  }\textbf {\bibinfo {volume} {478}},\ \bibinfo {pages} {89} (\bibinfo {year}
  {2011})},\ \Eprint {http://arxiv.org/abs/1106.3614} {arXiv:1106.3614}
  \BibitemShut {NoStop}\bibitem [{\citenamefont {Rivi{\`{e}}re}\ \emph {et~al.}(2011)\citenamefont
  {Rivi{\`{e}}re}, \citenamefont {Del{\'{e}}glise}, \citenamefont {Weis},
  \citenamefont {Gavartin}, \citenamefont {Arcizet}, \citenamefont
  {Schliesser},\ and\ \citenamefont {Kippenberg}}]{Riviere2011}  \BibitemOpen
  \bibfield  {author} {\bibinfo {author} {\bibfnamefont {R.}~\bibnamefont
  {Rivi{\`{e}}re}}, \bibinfo {author} {\bibfnamefont {S.}~\bibnamefont
  {Del{\'{e}}glise}}, \bibinfo {author} {\bibfnamefont {S.}~\bibnamefont
  {Weis}}, \bibinfo {author} {\bibfnamefont {E.}~\bibnamefont {Gavartin}},
  \bibinfo {author} {\bibfnamefont {O.}~\bibnamefont {Arcizet}}, \bibinfo
  {author} {\bibfnamefont {A.}~\bibnamefont {Schliesser}}, \ and\ \bibinfo
  {author} {\bibfnamefont {T.~J.}\ \bibnamefont {Kippenberg}},\ }\href
  {\doibase 10.1103/PhysRevA.83.063835} {\bibfield  {journal} {\bibinfo
  {journal} {Physical Review A}\ }\textbf {\bibinfo {volume} {83}},\ \bibinfo
  {pages} {063835} (\bibinfo {year} {2011})},\ \Eprint
  {http://arxiv.org/abs/1011.0290} {arXiv:1011.0290} \BibitemShut {NoStop}\end{thebibliography}

\begin{thebibliography}{10}\makeatletter
\providecommand \@ifxundefined [1]{ \@ifx{#1\undefined}
}\providecommand \@ifnum [1]{ \ifnum #1\expandafter \@firstoftwo
 \else \expandafter \@secondoftwo
 \fi
}\providecommand \@ifx [1]{ \ifx #1\expandafter \@firstoftwo
 \else \expandafter \@secondoftwo
 \fi
}\providecommand \natexlab [1]{#1}\providecommand \enquote  [1]{``#1''}\providecommand \bibnamefont  [1]{#1}\providecommand \bibfnamefont [1]{#1}\providecommand \citenamefont [1]{#1}\providecommand \href@noop [0]{\@secondoftwo}\providecommand \href [0]{\begingroup \@sanitize@url \@href}\providecommand \@href[1]{\@@startlink{#1}\@@href}\providecommand \@@href[1]{\endgroup#1\@@endlink}\providecommand \@sanitize@url [0]{\catcode `\\12\catcode `\$12\catcode
  `\&12\catcode `\#12\catcode `\^12\catcode `\_12\catcode `\%12\relax}\providecommand \@@startlink[1]{}\providecommand \@@endlink[0]{}\providecommand \url  [0]{\begingroup\@sanitize@url \@url }\providecommand \@url [1]{\endgroup\@href {#1}{\urlprefix }}\providecommand \urlprefix  [0]{URL }\providecommand \Eprint [0]{\href }\providecommand \doibase [0]{http://dx.doi.org/}\providecommand \selectlanguage [0]{\@gobble}\providecommand \bibinfo  [0]{\@secondoftwo}\providecommand \bibfield  [0]{\@secondoftwo}\providecommand \translation [1]{[#1]}\providecommand \BibitemOpen [0]{}\providecommand \bibitemStop [0]{}\providecommand \bibitemNoStop [0]{.\EOS\space}\providecommand \EOS [0]{\spacefactor3000\relax}\providecommand \BibitemShut  [1]{\csname bibitem#1\endcsname}\let\auto@bib@innerbib\@empty
\bibitem [{\citenamefont {Gardiner}\ and\ \citenamefont
  {Zoller}(2004)}]{Sgardiner2004quantum}  \BibitemOpen
  \bibfield  {author} {\bibinfo {author} {\bibfnamefont {C.}~\bibnamefont
  {Gardiner}}\ and\ \bibinfo {author} {\bibfnamefont {P.}~\bibnamefont
  {Zoller}},\ }\href {https://books.google.de/books?id=a{\_}xsT8oGhdgC} {\emph
  {\bibinfo {title} {{Quantum Noise: A Handbook of Markovian and Non-Markovian
  Quantum Stochastic Methods with Applications to Quantum Optics}}}},\ Springer
  Series in Synergetics\ (\bibinfo  {publisher} {Springer},\ \bibinfo {year}
  {2004})\BibitemShut {NoStop}\bibitem [{\citenamefont {Aspelmeyer}\ \emph {et~al.}(2014)\citenamefont
  {Aspelmeyer}, \citenamefont {Kippenberg},\ and\ \citenamefont
  {Marquardt}}]{SAspelmeyer2014}  \BibitemOpen
  \bibfield  {author} {\bibinfo {author} {\bibfnamefont {M.}~\bibnamefont
  {Aspelmeyer}}, \bibinfo {author} {\bibfnamefont {T.~J.}\ \bibnamefont
  {Kippenberg}}, \ and\ \bibinfo {author} {\bibfnamefont {F.}~\bibnamefont
  {Marquardt}},\ }\href {\doibase 10.1103/RevModPhys.86.1391} {\bibfield
  {journal} {\bibinfo  {journal} {Reviews of Modern Physics}\ }\textbf
  {\bibinfo {volume} {86}},\ \bibinfo {pages} {1391} (\bibinfo {year}
  {2014})}\BibitemShut {NoStop}\bibitem [{\citenamefont {Gardiner}\ and\ \citenamefont
  {Collett}(1985)}]{SGardiner1985}  \BibitemOpen
  \bibfield  {author} {\bibinfo {author} {\bibfnamefont {C.~W.}\ \bibnamefont
  {Gardiner}}\ and\ \bibinfo {author} {\bibfnamefont {M.~J.}\ \bibnamefont
  {Collett}},\ }\href {\doibase 10.1103/PhysRevA.31.3761} {\bibfield  {journal}
  {\bibinfo  {journal} {Physical Review A}\ }\textbf {\bibinfo {volume} {31}},\
  \bibinfo {pages} {3761} (\bibinfo {year} {1985})}\BibitemShut {NoStop}\bibitem [{\citenamefont {Peterson}\ \emph {et~al.}(2017)\citenamefont
  {Peterson}, \citenamefont {Lecocq}, \citenamefont {Cicak}, \citenamefont
  {Simmonds}, \citenamefont {Aumentado},\ and\ \citenamefont
  {Teufel}}]{SPeterson2017}  \BibitemOpen
  \bibfield  {author} {\bibinfo {author} {\bibfnamefont {G.~A.}\ \bibnamefont
  {Peterson}}, \bibinfo {author} {\bibfnamefont {F.}~\bibnamefont {Lecocq}},
  \bibinfo {author} {\bibfnamefont {K.}~\bibnamefont {Cicak}}, \bibinfo
  {author} {\bibfnamefont {R.~W.}\ \bibnamefont {Simmonds}}, \bibinfo {author}
  {\bibfnamefont {J.}~\bibnamefont {Aumentado}}, \ and\ \bibinfo {author}
  {\bibfnamefont {J.~D.}\ \bibnamefont {Teufel}},\ }\href {\doibase
  10.1103/PhysRevX.7.031001} {\bibfield  {journal} {\bibinfo  {journal}
  {Physical Review X}\ }\textbf {\bibinfo {volume} {7}},\ \bibinfo {pages}
  {031001} (\bibinfo {year} {2017})},\ \Eprint
  {http://arxiv.org/abs/1703.05269} {arXiv:1703.05269} \BibitemShut {NoStop}\bibitem [{\citenamefont {Bernier}\ \emph {et~al.}(2017)\citenamefont
  {Bernier}, \citenamefont {T{\'{o}}th}, \citenamefont {Koottandavida},
  \citenamefont {Ioannou}, \citenamefont {Malz}, \citenamefont {Nunnenkamp},
  \citenamefont {Feofanov},\ and\ \citenamefont {Kippenberg}}]{SBernier2017}  \BibitemOpen
  \bibfield  {author} {\bibinfo {author} {\bibfnamefont {N.~R.}\ \bibnamefont
  {Bernier}}, \bibinfo {author} {\bibfnamefont {L.~D.}\ \bibnamefont
  {T{\'{o}}th}}, \bibinfo {author} {\bibfnamefont {A.}~\bibnamefont
  {Koottandavida}}, \bibinfo {author} {\bibfnamefont {M.~A.}\ \bibnamefont
  {Ioannou}}, \bibinfo {author} {\bibfnamefont {D.}~\bibnamefont {Malz}},
  \bibinfo {author} {\bibfnamefont {A.}~\bibnamefont {Nunnenkamp}}, \bibinfo
  {author} {\bibfnamefont {A.~K.}\ \bibnamefont {Feofanov}}, \ and\ \bibinfo
  {author} {\bibfnamefont {T.~J.}\ \bibnamefont {Kippenberg}},\ }\href
  {\doibase 10.1038/s41467-017-00447-1} {\bibfield  {journal} {\bibinfo
  {journal} {Nature Communications}\ }\textbf {\bibinfo {volume} {8}},\
  \bibinfo {pages} {604} (\bibinfo {year} {2017})},\ \Eprint
  {http://arxiv.org/abs/1612.08223} {arXiv:1612.08223} \BibitemShut {NoStop}\bibitem [{\citenamefont {Woolley}\ and\ \citenamefont
  {Clerk}(2013)}]{SWoolley2013}  \BibitemOpen
  \bibfield  {author} {\bibinfo {author} {\bibfnamefont {M.~J.}\ \bibnamefont
  {Woolley}}\ and\ \bibinfo {author} {\bibfnamefont {A.~A.}\ \bibnamefont
  {Clerk}},\ }\href {\doibase 10.1103/PhysRevA.87.063846} {\bibfield  {journal}
  {\bibinfo  {journal} {Physical Review A}\ }\textbf {\bibinfo {volume} {87}},\
  \bibinfo {pages} {063846} (\bibinfo {year} {2013})},\ \Eprint
  {http://arxiv.org/abs/1304.4059} {arXiv:1304.4059} \BibitemShut {NoStop}\bibitem [{\citenamefont {Malz}\ and\ \citenamefont
  {Nunnenkamp}(2016{\natexlab{a}})}]{SMalz2016}  \BibitemOpen
  \bibfield  {author} {\bibinfo {author} {\bibfnamefont {D.}~\bibnamefont
  {Malz}}\ and\ \bibinfo {author} {\bibfnamefont {A.}~\bibnamefont
  {Nunnenkamp}},\ }\href {\doibase 10.1103/PhysRevA.94.023803} {\bibfield
  {journal} {\bibinfo  {journal} {Physical Review A}\ }\textbf {\bibinfo
  {volume} {94}},\ \bibinfo {pages} {023803} (\bibinfo {year}
  {2016}{\natexlab{a}})},\ \Eprint {http://arxiv.org/abs/1605.04749}
  {arXiv:1605.04749} \BibitemShut {NoStop}\bibitem [{\citenamefont {Malz}\ and\ \citenamefont
  {Nunnenkamp}(2016{\natexlab{b}})}]{SMalz2016b}  \BibitemOpen
  \bibfield  {author} {\bibinfo {author} {\bibfnamefont {D.}~\bibnamefont
  {Malz}}\ and\ \bibinfo {author} {\bibfnamefont {A.}~\bibnamefont
  {Nunnenkamp}},\ }\href {\doibase 10.1103/PhysRevA.94.053820} {\bibfield
  {journal} {\bibinfo  {journal} {Physical Review A}\ }\textbf {\bibinfo
  {volume} {94}},\ \bibinfo {pages} {053820} (\bibinfo {year}
  {2016}{\natexlab{b}})},\ \Eprint {http://arxiv.org/abs/1610.00154}
  {arXiv:1610.00154} \BibitemShut {NoStop}\bibitem [{\citenamefont {Li}\ \emph {et~al.}(2015)\citenamefont {Li},
  \citenamefont {Haghighi}, \citenamefont {Malossi}, \citenamefont {Zippilli},\
  and\ \citenamefont {Vitali}}]{SLi2015}  \BibitemOpen
  \bibfield  {author} {\bibinfo {author} {\bibfnamefont {J.}~\bibnamefont
  {Li}}, \bibinfo {author} {\bibfnamefont {I.~M.}\ \bibnamefont {Haghighi}},
  \bibinfo {author} {\bibfnamefont {N.}~\bibnamefont {Malossi}}, \bibinfo
  {author} {\bibfnamefont {S.}~\bibnamefont {Zippilli}}, \ and\ \bibinfo
  {author} {\bibfnamefont {D.}~\bibnamefont {Vitali}},\ }\href {\doibase
  10.1088/1367-2630/17/10/103037} {\bibfield  {journal} {\bibinfo  {journal}
  {New Journal of Physics}\ }\textbf {\bibinfo {volume} {17}},\ \bibinfo
  {pages} {103037} (\bibinfo {year} {2015})}\BibitemShut {NoStop}\bibitem [{\citenamefont {Metelmann}\ and\ \citenamefont
  {Clerk}(2015)}]{SMetelmann2015}  \BibitemOpen
  \bibfield  {author} {\bibinfo {author} {\bibfnamefont {A.}~\bibnamefont
  {Metelmann}}\ and\ \bibinfo {author} {\bibfnamefont {A.~A.}\ \bibnamefont
  {Clerk}},\ }\href {\doibase 10.1103/PhysRevX.5.021025} {\bibfield  {journal}
  {\bibinfo  {journal} {Physical Review X}\ }\textbf {\bibinfo {volume} {5}},\
  \bibinfo {pages} {021025} (\bibinfo {year} {2015})},\ \Eprint
  {http://arxiv.org/abs/1502.07274} {arXiv:1502.07274} \BibitemShut {NoStop}\end{thebibliography}
  \end{widetext}
\end{document}